\documentclass{aa} 
\usepackage{epsfig,graphicx,natbib,txfonts,url}
\usepackage[usenames]{color}
\bibpunct{(}{)}{;}{a}{}{,}    %% A&A fix to natbib
\def\cite#1{\citealp{#1}}     %% restore old astroncite \cite command

\def\figspath{.}
\newcount\longrefs
\def\aap{\ifnum\longrefs=1 {Astron.\ Astrophys.}\else 
                           {A\hbox{\rm \&}A}\fi}
\def\aapr{\ifnum\longrefs=1 {Astron.\ Astrophys.\ Rev.}\else 
                            {A\hbox{\rm \&}AR}\fi}
\def\aaps{\ifnum\longrefs=1 {Astron.\ Astrophys.\ Suppl.}\else 
                            {A\hbox{\rm \&}A Suppl.}\fi}
\def\aj{\ifnum\longrefs=1 {Astron.\ J.}\else 
                          {AJ}\fi} 
\def\ao{\ifnum\longrefs=1 {Applied Optics}\else 
                           {Appl.\ Opt.}\fi} 
\def\aspcs{\ifnum\longrefs=1 {Astron.\ Soc.\ Pacific Conf. Series}\else 
                           {ASP Conf.\ Ser.}\fi} 
\def\apj{\ifnum\longrefs=1 {Astrophys.\ J.}\else 
                           {ApJ}\fi} 
\def\apjl{\ifnum\longrefs=1 {Astrophys.\ J. Lett.}\else 
                            {ApJ}\fi} 
\def\aplett{\ifnum\longrefs=1 {Astrophys.\ J. Lett.}\else 
                            {ApJ}\fi} 
\def\apjs{\ifnum\longrefs=1 {Astrophys.\ J. Suppl.}\else 
                            {ApJS}\fi}
\def\apss{\ifnum\longrefs=1 {Astrophys.\ and Space Science}\else 
                            {Astrophys.\ Space Sci.}\fi}
\def\araa{\ifnum\longrefs=1 {Ann.\ Rev.\ Astron.\ Astrophys.}\else 
                            {ARA\hbox{\rm \&}A}\fi}
\def\azh{\ifnum\longrefs=1 {Astronomicheskii Zhurnal}\else 
                            {Astron.\ Zhur.}\fi}
\def\baas{\ifnum\longrefs=1 {Bull.\ Am.\ Astron.\ Soc.}\else 
                            {BAAS}\fi}
\def\bain{\ifnum\longrefs=1 {Bull.\ Astronom.\ Institutes Netherlands}\else
                            {Bull.\ Astr.\ Inst.\ Neth.}\fi}
\def\gca{\ifnum\longrefs=1 {Geochim.\ Cosmochim.\ Acta}\else 
                           {Geochim.\ Cosmochim.\ Acta}\fi}
\def\grl{\ifnum\longrefs=1 {Geophys.\ Res.\ Lett.}\else 
                           {Geoph.\ Res.\ Lett.}\fi}
\def\iaucirc{\ifnum\longrefs=1 {IAU Circulars}\else 
                          {IAU Circ.}\fi}
\def\ip{\ifnum\longrefs=1 {in press}\else 
                          {in press}\fi}
\def\jgr{\ifnum\longrefs=1 {J.\ Geophys.\ Res.}\else 
                           {J.\ Geophys.\ Res.}\fi}  
\def\jrasc{\ifnum\longrefs=1 {J.\ Royal Astron.\ Soc.\ Canada}\else 
                           {JRAS Can.}\fi}  
\def\mnras{\ifnum\longrefs=1 {Mon.\ Not.\ Roy.\ Astron.\ Soc.}\else 
                             {MNRAS}\fi} 
\def\nat{\ifnum\longrefs=1 {Nature}\else 
                           {Nat}\fi}
\def\pasj{\ifnum\longrefs=1 {Pub.\ Astron.\ Soc.\ Japan}\else 
                            {PASJ}\fi} 
\def\pasp{\ifnum\longrefs=1 {Pub.\ Astron.\ Soc.\ Pacific}\else 
                            {PASP}\fi} 
\def\physscr{\ifnum\longrefs=1 {Physica Scripta}\else 
                            {Phys.\ Scrip.}\fi} 
\def\planss{\ifnum\longrefs=1 {Planetary \& Space Science}\else 
                            {Plan. \& Space Sci.}\fi} 
\def\procspie{\ifnum\longrefs=1 {Proc.\ SPIE}\else 
                            {Proc.\ SPIE}\fi} 
\def\qjras{\ifnum\longrefs=1 {Quarterly J.\ Royal Astron.\ Soc.}\else 
                            {QJRAS}\fi} 
\def\sa{\ifnum\longrefs=1 {Soviet Astron..}\else 
                               {Sov.\ Astron.}\fi}
\def\skytel{\ifnum\longrefs=1 {Sky \& Telescope}\else 
                            {Sky \& Tel.}\fi} 
\def\solphys{\ifnum\longrefs=1 {Solar Phys.}\else 
                               {Sol.\ Phys.}\fi}
\def\ssr{\ifnum\longrefs=1 {Space Science Rev.}\else 
                               {Space\ Sci.\ Rev.}\fi}

\def\nl{,\ } %%\def\nl{\newline}  %% redefine as \newline for mail addresses

\def\ITA{Institute of Theoretical Astrophysics\nl
         University of Oslo\nl
         P.O. Box 1029, Blindern\nl N--0315 Oslo\nl Norway}

\def\SIU{Sterrekundig Instituut\nl Utrecht University\nl Postbus 80\,000\nl
         NL--3508 TA Utrecht\nl The Netherlands}
\def\SPO{National Solar Observatory/Sacramento Peak\nl P.O. Box 62\nl 
         Sunspot, NM 88349--0062\nl USA}

\def\rmit#1{{\it #1}}              %% italics (RR style, Kluwer)
\def\etal{\rmit{et al.}}           %% use \etal\ for space behind it        
           
\def\ie{\rmit{i.e.,}}              %% , required (Webster 1681)
\def\eg{\rmit{e.g.,}}              %% , required (Webster 1681)
\def\specchar#1{\uppercase{#1}}    %% to be redefined for A&A, small caps
\def\CaII{\mbox{Ca\,\specchar{ii}}}

\def\FeI{\mbox{Fe\,\specchar{i}}}

\def\Hmin{\hbox{\rmH$^{^{_-}}\!$}}      %% H^min, very elegant
\def\MgI{\mbox{Mg\,\specchar{i}}} 
\def\MgII{\mbox{Mg\,\specchar{ii}}} 
 
\def\MnI{\mbox{Mn\,\specchar{i}}}

\def\Halpha{\mbox{H\hspace{0.1ex}$\alpha$}} %% \Halpha\ for space behind it

\def\CaIIK{\mbox{Ca\,\specchar{ii}\,\,K}}       %% use \CaIIK\ for space

\def\CaIIHK{\mbox{Ca\,\specchar{ii}\,\,H\,\&\,K}}
\def\HK{\mbox{H\,\&\,K}}
\def\MgIIk{\mbox{Mg\,\specchar{ii}\,\,k}}

\def\MgIIhk{\mbox{Mg\,\specchar{ii}\,\,h\,\&\,k}}
\def\hk{\mbox{h\,\&\,k}}
\def\kthree{\mbox{k$_3$}}

\def\kone{\mbox{k$_1$}}

 \def\rmH{{\rm H}}

\def\rml{{\rm l}}

\def\kms{\hbox{km$\;$s$^{-1}$}}

\def\is{\!=\!}                             %% = in text for tighter spacing
\def\gf{\mbox{$g \! f$}}                   %% \gf\ for space behind it

\def\={\hbox{$\!=\!$}}                     %% less space around =

\def\Mnline{Mn\,\specchar{i} 5394.7~\AA} 
\def\Feline{Fe\,\specchar{i} 5395.2~\AA} 

\def\rmit#1{#1}               %% A&A & ApJ: latin abbreviations in Roman
\def\specchar#1{{\sc #1}}     %% A&A style for ionization stages
\def\revpar{}                 %% ignore revision change markers
\long\def\rev#1{#1}           %% ignore revision change boldface

\begin{document} 

\title{Explanation of the activity sensitivity of Mn\,I\,5394.7\,\AA}
\titlerunning{Explanation of the activity sensitivity of \Mnline}

\author{N. Vitas \inst{1}       
        \and
        B. Viticchi\`e \inst{2}
        \and
        R.J. Rutten \inst{1,3,4}
        \and 
        A. V\"ogler \inst{1}
}

\authorrunning{N. Vitas \etal}

\institute{\SIU
           \and
           Dipartimento di Fisica, Universit\`a degli Studi di Roma
           ``Tor Vergata'', Via dell Ricerca Scientifica,
           I-00133 Roma, Italy
           \and \ITA
           \and \SPO
}

\date{Received; accepted}
\offprints{N. Vitas\\
           \email{N.Vitas@uu.nl}}

%%%%%%%%%%%%%%%%%%%%%%%%%%%%%%%%%%%%%%%%%%%%%%%%%%%%%%%%%%%%%%%%%% ABSTRACT
\abstract  
% context
  {There is a \rev{long-standing controversy concerning the reason
  why} the \Mnline\ line in the solar irradiance spectrum brightens
  more at larger activity than most other photospheric lines.  The
  claim that this \rev{activity sensitivity is caused by} spectral
  interlocking to chromospheric emission in \MgII\ \hk\ is disputed.}
%% aims
  {To settle and close this debate.}
%% Methods: 
  {Classical one-dimensional modeling is used for demonstration;
   modern three-dimensional MHD simulation for verification and analysis.}
%% Results:
  {The \Mnline\ line thanks its unusual sensitivity to solar activity
  to its hyperfine structure.  This overrides the thermal and granular
  Doppler smearing through which the other, narrower, photospheric
  lines lose such sensitivity.  We take the nearby \Feline\ line as
  example of the latter and analyze the formation of both lines in
  detail to demonstrate \rev{and explain} granular Doppler
  brightening.  \rev{We show that this affects all narrow lines.}
  Neither the chromosphere nor \MgII\ \hk\ play a role,
  \rev{nor is it correct to describe the activity sensitivity of
  \Mnline\ through plage models with outward increasing temperature
  contrast.}}
%% Conclusions:
  {The \Mnline\ line represents a proxy diagnostic of strong-field
  magnetic concentrations in the deep solar photosphere comparable to
  the G band and the blue wing of \Halpha, but not a better one than
  these.  The \MnI\ lines are more promising as diagnostic of weak
  fields in high-resolution Stokes polarimetry.}


\maketitle

%%%%%%%%%%%%%%%%%%%%%%%%%%%%%%%%%%%%%%%%%%%%%%%%%%%%%%%%%%%%%%%%%%%%%%%%%%%%
\section{Introduction}                              \label{sec:introduction}
%%%%%%%%%%%%%%%%%%%%%%%%%%%%%%%%%%%%%%%%%%%%%%%%%%%%%%%%%%%%%%%%%%%%%%%%%%%%

In this paper we analyze the formation of the solar \Mnline\ line in
the context of its global sensitivity to solar activity, a subject
which has received considerable attention based on the extensive
observations of this line by Livingston and coworkers and Vince and
coworkers
 (\cite{1987ApJ...314..808L}; %C Livingston++ luminosity variation 
  \cite{1998IAUS..185..459V}; %C Vince+Erkapic: chromospheric behavior
  Danilovic \& Vince 2004, 2005;
      \nocite{2004SerAJ.169...47D} %C Danilovic+Vince: Belgrade = KPNO
      \nocite{2005MmSAI..76..949D} %C Danilovic+Vince: correlation plage
  \cite{2004IAUS..223..645M}; %C Malanushenko++ imaging in MnI
  \cite{2005SerAJ.170...79D}; %C Danilovic++ time series MnI 
  Vince \etal\ 2005a, 2005b;
      \nocite{2005SerAJ.170..115V} %C Vince++ MnI in plages
      \nocite{2005SoPh..229..273V} %C Vince++ MnI in ARs
  \cite{2007ApJ...657.1137L}). %C Livingston++ flux cycle variations

In hindsight, the starting publications were by
  \citet{1937ApJ....86..499T} %C Thackeray MnI-MgII coupling Mira's
who pointed out that the violet wing of \MgIIk\ (line center \kthree\ at
2795.53~\AA) overlaps with
\MnI\,2794.82~\AA\ and so may produce optical pumping of
that and other \MnI\ lines in stellar spectra,
and by
  \citet{1952ApJ...115..199A} %C Abt HFS  # Athur = Helmut!
who pointed out that the unusual widths of the \MnI\ lines in the
solar spectrum are due to hyperfine structure.  These two points are
key ones in almost all later work on solar \MnI\ lines, effectively
dividing this into two categories.  The first addresses the unusual
sensitivity of the \Mnline\ line to global solar activity, mostly
debating the \revpar forceful claim by
  \citet{2001A&A...369L..13D} %C Doyle++ solar MnI variation explained
that this is explained by Thackeray's \MgIIk\ coincidence operating in
the solar chromosphere.  This activity response is also our subject
here, but we establish and explain \revpar that neither \MgIIk\ nor
the chromosphere has anything to do with it.  

We first summarize the observations and discuss these conflicting
views in the next paragraphs of this introduction, and we then put the
issue to rest by demonstrating that Abt's hyperfine structure is the
key agent through reducing spectral-line sensitivity to thermal and
convective Dopplershifts outside magnetic concentrations.

The second category of solar \MnI\ papers also addresses usage of
\MnI\ lines as diagnostic of solar magnetism but concentrates on 
quantitative measurement of weak internetwork fields exploiting
the intricate line-center opacity variation with wavelength
imposed by hyperfine structure to disentangle weak- and strong-field
signatures in Stokes polarimetry
  (L\'opez Ariste \etal\ 2002, 2006a, 2006b;
  \nocite{2002ApJ...580..519L} %C LopezAriste++: HFS as B diagnostic
  \nocite{2006A&A...454..663L} %C LopezAriste++ QS B with Mn
  \nocite{2006ASPC..358...54L} %C LopezAriste++ QS B with B: km size
  \cite{2007ApJ...659..829A}; %C AsensioRamos++ near-IR Mn weak field
%%  \cite{2007MmSAI..78...54R}; %? Ramirez+LopezAriste MnI553
    \cite{2008ApJ...675..906S}). %C Sanchez++: QS B measurement from HFS
This potentially more fruitful topic is not addressed here.

%%%%%%%%%%%%%%%%%%%%%%%%%%%%%%%%%%%%%%%%%%%%%%%%%%%%%%%%%%%%%%%%%%%%%%%%%%%%
\subsection{Activity modulation}
%%%%%%%%%%%%%%%%%%%%%%%%%%%%%%%%%%%%%%%%%%%%%%%%%%%%%%%%%%%%%%%%%%%%%%%%%%%%
Livingston's inclusion of the \Mnline\ line into his long-term
full-disk ``sun-as-a-star'' line profile monitoring from 1979 onwards
was suggested by Elste who had pointed out that their large hyperfine
structure makes the \MnI\ lines less sensitive to the questionable
microturbulence parameter than other ground-state neutral-metal lines
that may serve as temperature diagnostic
  (\cite{1978SoPh...59..275E}; %C Elste+Teske: sensitivity network
   \cite{1987SoPh..107...47E}). %C Elste temperature diagnostic = similar
\rev{Livingston found} that this line is the only photospheric line in 
his full-disk monitoring that shows appreciable variation with the
cycle, in good concert with the \CaIIK\ full-disk intensity variation.
Its equivalent width in the irradiance spectrum varies by
up to 2\%
  (\cite{1987ApJ...314..808L}). %C Livingston++ luminosity variation 
Figure~16 in the overview paper of
  \citet{2007ApJ...657.1137L} %C Livingston++ flux cycle variations
displays the variation in \rev{relative line depth (\ie\ the minimum
intensity of the line in the full-disk spectrum expressed as fraction
of the continuum intensity and measured from the latter, plotted}
upside down and therefore labeled ``central intensity'' in the caption).
\revpar
%% during the two monitoring periods without problematic instrument changes.
The same data are plotted as \rev{relative} line depth in Fig.\,2 of
  \citet{2007msfa.conf..189D}, %C Danilovic++ magnetic source MnI
overlaid by a theoretical modeling curve.  \rev{The} relative
line-center intensity increases about 2\% from cycle minimum to
maximum, slightly more than the corresponding decrease in equivalent
width.

  \citet{2005SerAJ.170..115V} %C Vince++ MnI in plages
used observations at the Crimea Observatory including Zeeman
polarimetry to measure the changes in \Mnline\ between plages with
different apparent magnetic flux density.  They found that the line
weakens with increasing flux, as concluded already by
  \citet{1978SoPh...59..275E} %C Elste+Teske: sensitivity network
  and
  \citet{1987SoPh..107...47E}. %C Elste temperature diagnostic 
\revpar
%% The full width at half maximum of the line-depth profile does not
%% change much, so that the  equivalent width weakening tracks
%% the line-center brightening.

%%%%%%%%%%%%%%%%%%%%%%%%%%%%%%%%%%%%%%%%%%%%%%%%%%%%%%%%%%%%%%%%%%%%%%%%%%%%
\subsection{Chromospheric interpretation}
%%%%%%%%%%%%%%%%%%%%%%%%%%%%%%%%%%%%%%%%%%%%%%%%%%%%%%%%%%%%%%%%%%%%%%%%%%%%
%
  \citet{2001A&A...369L..13D} %C Doyle++ solar MnI variation explained
gave their paper the title ``{\em Solar \MnI\ 5432/5395~\AA\ line
formation explained}'' which we paraphrase in our title above.
They used 
\revpar
%% detailed multi-species multi-level 
NLTE computations to \rev{claim} that these
\MnI\ lines are sensitive to optical pumping through the overlap
coincidence with \MgIIk\ noted by Thackeray.  The \rev{claim}
consisted of displaying \MnI\ profiles for different solar atmosphere
models without and with taking \MgII\ \hk\ into account.  The different
models specified ad-hoc variation in the onset heights of the
chromospheric and transition-region temperature rises.  Appreciable
variation of the two \MnI\ lines was found and attributed to the 
\rev{spectral interlocking}.

However, closer inspection undermines this \revpar claim.  Large
changes, of order 30\% in line depth, were shown to occur when \MgII\
\hk\ and all other blanketing lines are not taken into account,
but this is not a relevant test since deletion of the ultraviolet line
haze invalidates the ionization equilibrium evaluation for any species
with intermediate ionization energy.  \revpar The same test would
be as dramatic for any optical \FeI\ line.  The changes in the profile
of \Mnline\ between the cases of \MnI\,--\,\MgII\ coupling and no coupling
were negligible unless the model possessed a very deep-lying
chromosphere and transition region \revpar producing unrealistic high peaks
in \CaII\ \HK\ and \MgII\ \hk.  Even then, the computed brightening of
\Mnline\ amounted to only a few percent, yet to be diluted through a
filling factor of order 0.01 to represent the contribution of
active-sun plage in full-disk averaging.

  \citet{2001A&A...369L..13D} %C Doyle++ solar MnI variation explained
added no further analysis (such as \rev{specification of} NLTE departures,
radiation fields, source functions) \revpar but only verbal
explanation, literally: ``{\em because of the huge absorption in
\MgII, the local continuum for the \MnI\ UV lines changes.  There is
less flux and thus fewer photons to be absorbed and hence the ground
level is consequently more populated\/}''.
\revpar
%% which suggests an optically-thin grasp of line formation.
What actually happens in such coupling is that the quasi-continuous
\MgII\ wing opacity increases the height of photon escape and enforces
LTE behavior to considerably larger height than the \MnI\ lines might
maintain on their own, up to the height where \MgIIk\ photon loss
causes source function \rev{departure from} the Planck function.
Thus, thanks to the wavelength coincidence, departures from LTE set in
only at exceptionally large height, and these \revpar anyhow
represent a much larger fractional population change for the upper
level than \revpar for the ground level, affecting the source function
much more than the opacity.

%RR not photon loss - in the far wing probably J>B and S>B in PRD
%RR although this is k1 so indeed S=J<B

Optical pumping, also claimed in the paper's verbal explanation, is
perhaps easier grasped.  Super-Planckian radiation in a pumped
transition may overpopulate its upper level and so induce
super-Planckian source function excess and apparent brightening in
subordinate lines from the same upper level, in this case \MnI\
multiplet 4, and perhaps also in other lines through upper-level
coupling.  Such pumping was earlier established for a variety of
emission lines in the extended wings of \CaIIHK\
  (\eg\
   \cite{1971A&A....10...64C}; % Canfield CeII
   \cite{1980A&AS...39..415R}; % Rutten+Stencel HK limb
   \cite{1980ApJ...241..374C}), % Cram+Rutten+Lites FeII 
but these arise through coupling to more deeply escaping super-Planckian
radiation outside the \HK\ wings, a wholly different mechanism.  No
such pumping affects the \MnI\ lines, nor did pumping operate in the
computation of
  \citet{2001A&A...369L..13D}; %C Doyle++ solar MnI variation explained
the slight brightening of \Mnline\ which they obtained for deep-lying
chromospheres was simply contributed through the outer tail of the
contribution function.

%RR their plot has the MnI UV line in emission on top of that wing!

Actually, the computation was intrinsically wrong because the
\MnI\ multiplet UV1 lines do not coincide with the \MgII\ \hk\ cores
but lie beyond $\Delta \lambda = 0.7$~\AA\ in their wings, of which
the intensity is considerably overestimated when assuming complete
redistribution instead of partially coherent scattering -- one of the
worst locations in the whole solar spectrum to make this mistake.  In
the computations deep onsets of the chromospheric temperature rise
resulted in bright extended \MgII\ \hk\ wings, but in reality the
independent radiation fields in the inner wings decouple from the
Planck function already in the photosphere
  (\cite{1974ApJ...192..769M}). %C Milkey+Mihalas MgII hk PRD
This error is obvious when comparing the computed profiles in Fig.\,3
of
  \citet{2001A&A...369L..13D} %C Doyle++ solar MnI variation explained
\revpar 
%% (of which the intensity units are wrong) 
with observed \MgII\ \hk\ profiles such as the pioneering ones by
  \citet{1973A&A....22...61L} %C Lemaire+Skumanich MgII hk obs
in which even the strongest plage emission shows deep \kone\ minima at
$\Delta \lambda = \pm 0.5$~\AA\ from line center. \revpar
In fact, the
\MnI\ 2794.82~\AA\ line is visible as an absorption dip within the
deep \MgII\ \kone\ minimum not only in the reference spectrum in
Fig.\,1 of
  \citet{1995A&A...295..517S} %C Staath+Lemaire MgII hk
but also in all \MgII\ \hk\ profiles displayed in their Fig.\,2, and it
remains located within the \kone\ dip even in all limb spectra in
their Fig.\,11, both inside and outside the limb.  The latter result
from summation of chromospheric \MgIIk\ emission along the line of
sight so that their peak widths represent a maximum.  Therefore,
everywhere across the solar surface \MnI\ 2794.82~\AA\ lies in the
deep \kone\ dip which has a sub-Planckian source function due to
coherent scattering and does not at all respond to chromospheric
activity.  

In summary, although both the \Mnline\ line and the peaks in \MgII\
\hk\ are observed to track solar activity, the blending of the \MnI\
UV1 lines into the opaque \MgII\ \hk\ wings does not imply a viable
causal relationship, nor was one proven by
  \citet{2001A&A...369L..13D}. %C Doyle++ solar MnI variation explained
The \MnI\ UV1 lines lie too far from the chromospheric \hk\ emission
peaks for any pumping by these.  In addition, the \MnI\ multiplet 1
lines are formed much deeper 
  (\cite{2007ASPC..368..543V}).  %C Vitas++ Coimbra 
%

%===========================================================================
%% Fig.\,\ref{fig:BPcartoon} 
%===========================================================================
\begin{figure}
  \centering \includegraphics[width=88mm]{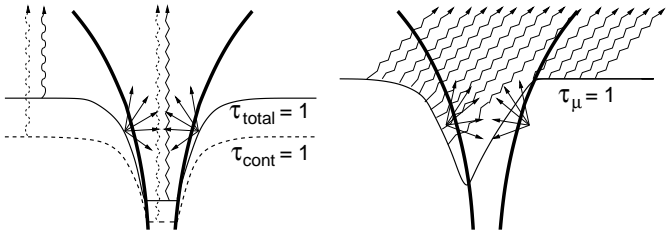}
  \caption[]{
  Schematic G-band bright-point enhancement in a magnetic concentration,
  portrayed here as a vertical cut through an idealized flaring
  fluxtube (thick curves) containing strong field, embedded in
  field-free quiet photosphere.  {\em Left\/}: in radial viewing the
  Wilson depression due to magnetic-pressure evacuation deepens the
  photon escape layer characterized as $\tau\is1$ surface to far below
  the outside surface.  The magnetic concentrations are much cooler there
  than the subsurface surroundings due to suppressed convection, but
  hotter than the outside photon escape layer due to the large depth
  and hot-wall irradiation, and have flatter temperature gradients.
  The correspondingly larger degree of dissociation of the CH
  molecules that make up the G band cause a yet larger effective
  Wilson depression in this spectral feature, producing larger
  brightness enhancement than in the continuum.  {\em Right\/}: In
  near-limb viewing the same lack of opacity along the slanted line of
  sight causes deeper, facular ``bright-stalk'' sampling of hot
  granules behind magnetic concentrations.  From
  \citet{1999ASPC..184..181R}; % Rutten Potsdam review 
  see also Fig.\,1 of 
  \citet{1976SoPh...50..269S}. % Spruit fluxtubes
}
\label{fig:BPcartoon}
\end{figure}
%===========================================================================
%RR I lessened in-tube difference at left to mimick Carlsson++ diagram
%RR flatter gradient: from Leenaarts BP-Ha; tau=1 has similar density

%RR does low density count?  Mats says so (dissociation) but it is
%RR accounted for by Wilson depression: around tau=1 same density as outside
%RR Ha wing also has same damping at same density?  Hard to say, need
%RR gas density panel in bottom row Fig4 Leenaarts Ha-BP.
%RR but flat extended T gradient implies steeper Pg gradient so indeed
%RR small damping along contribution function.
%RR asked Jorrit; result: indeed extended 

%%%%%%%%%%%%%%%%%%%%%%%%%%%%%%%%%%%%%%%%%%%%%%%%%%%%%%%%%%%%%%%%%%%%%%%%%%%%
\subsection{Photospheric interpretation}
%%%%%%%%%%%%%%%%%%%%%%%%%%%%%%%%%%%%%%%%%%%%%%%%%%%%%%%%%%%%%%%%%%%%%%%%%%%%
Height of formation estimates for the optical \MnI\ lines
based on standard modeling suggests that they are purely photospheric
 (\cite{1989KiIND.........G}; %T Gurtovenko+Kostik book
  \cite{2005MSAIS...7..164V}). %C Vitas heights of formation 
Observational evidence that \Mnline\ is purely photospheric has been
collected by
  \citet{2005SoPh..229..273V} %C Vince++ MnI in ARs
who showed that \Mnline\ bisectors show characteristic 
photospheric shapes and center-limb behavior,
and by
  \citet{2004IAUS..223..645M} %C Malanushenko++ imaging in MnI
who compared spectroheliogram scans in \Mnline\ and the nearby \MnI\
5420.4~\AA\ line with other lines 
%RR from \FeI\ and \SiI\ 
and a magnetogram.  The \MnI\ lines show network bright, closely
mimicking the unsigned magnetogram signal which is purely photospheric.

%% To their surprise, the network also brightens in \SiI\ 5421.2~\AA.

%RR the BaII line doesn't sense thermal and has some HFS - that it?

%RR Olena/Elena Malanushenko gave me these spectrum scans in June.
%RR I bet the calibration was done by giving granulation the
%RR same grayscale in all her panels - they look like that.
%RR all these ``images'' should instead be on abs int greyscale, like
%RR our MURaM ones.  Presumably the MnI granulation then again darkens.
%RR Same scatter plots as in our paper would be good demonstration.  
%RR Nice for some conference poster publication (Bangalore?)

The arguments \rev{above} against a chromospheric interpretation and
these observational indications of photospheric formation together
suggest that the propensity of \MnI\ lines to track solar activity in
their line-center brightness may be akin to the contrast brightening
that network and plage show in the G band.  We therefore summarize
G-band bright-point formation, where ``bright points'' stands for
kilo-Gauss magnetic concentrations.  Their enhanced photospheric
brightness in continuum intensity and further contrast increase in
G-band imaging has been studied in extenso and is well understood,
both for their on-disk appearance as filigree and near-limb appearance
as faculae.  A brief review with the key references is given in the
introduction of
  \citet{2005A&A...441.1183D}; % de Wijn++ patches tomo4
the cartoon in Fig.\,\ref{fig:BPcartoon} schematizes 
\revpar the magnetostatic fluxtube paradigm of
  \citet{1967SoPh....1..478Z} %C Zwaan: invisible sunspots and
and
  \citet{1976SoPh...50..269S}. %? Spruit, magnetostatic fluxtube 
The latest numerical verifications of this concept are the MHD
simulations of
  \citet{2004ApJ...607L..59K}, % Keller++: faculae
  \citet{2004ApJ...610L.137C}, %C Carlsson++ Gband BPs
and  
  \citet{2004A&A...427..335S}. % Shelyag++ MURaM BPs
The upshot is that the G band shows magnetic concentrations with enhanced
contrast because its considerable line opacity lessens through
molecular dissociation within the concentrations while its LTE formation
implies good temperature mapping.  Similarly, the extended blue wing
of \Halpha\ brightens also in magnetic concentrations through lessening of
collisional broadening plus LTE formation
  (\cite{2006A&A...449.1209L}). %C BPs blue wing Ha

Hence, for manganese we seek a property that enhances the reduction of
line opacity in fluxtubes over that in comparable lines say from \FeI,
enhancing the corresponding ``line gap'' phenomenon.  A first
consideration is that \Mnline\ is relatively sensitive to
temperature, as pointed out already by
  \citet{1978SoPh...59..275E}, %C Elste+Teske: sensitivity network
because it originates from a neutral-metal ground state.  Such lines
do not suffer from the cancellation in LTE response to temperature
increase that lines from excited levels have through 
\rev{masking of
   higher-temperature perturbations by radially moving the formation of the
   line outward where it samples lower temperatures}
  (see Fig.\,4 of
   \cite{2005ESASP.596E..15L}). %C Leenaarts++ Lindau
In addition, \Mnline\ is also a somewhat forbidden intersystem
transition which makes its source function obey LTE more closely than
for higher-probability lines.  However, both these properties are
unlikely to play an important role since
  \citet{2004IAUS..223..645M} %C Malanushenko++ imaging in MnI
found that \MnI\ 5420.4~\AA, a member of multiplet 4 at 2.14~eV
excitation, brightens about as much in network.

The obvious remaining property which makes \MnI\ lines differ from
others is their large hyperfine structure. How can this cause unusual
brightness enhancement in strong-field magnetic concentrations?  
  \citet{1978SoPh...59..275E} %C Elste+Teske: sensitivity network
pointed out that it lessens sensitivity to the ``turbulence'' that was
needed to explain other lines in classical one-dimensional modeling of
the spatially-averaged solar spectrum.  The ill-famous
``microturbulence'' and ``macroturbulence'' parameters were supposed to
emulate the reality of convective and oscillatory inhomogeneities
which
\revpar
%% affect Fraunhofer lines and 
make a solar-atlas line profile represent a spatio-temporal
average over widely fluctuating and Dopplershifted instantaneous
local profiles.  Lines that should be deep and narrow are so smeared
into shallower average depressions.  However, lines that are already wide
intrinsically through hyperfine broadening suffer less shallowing by
being less sensitive to Dopplershifts.  The culprit may therefore not
be the hyperfine structure of \MnI\ lines but rather the heavy
hydrodynamic smearing of all the other, narrow lines that occurs in
the granulation outside magnetic concentrations.  We test this idea
below and find it is correct, turning the analysis into \rev{one} 
of general \FeI\ line formation rather than specific \MnI\ line
formation.

%RR in fluxtubes as much Doppler but there the lines vanish anyhow

We first demonstrate this idea with classical one-dimensional modeling
in Sect.~\ref{sec:onedimdemo}, then verify it through
three-dimensional MHD simulation in Sect.~\ref{sec:simulation},
\rev{and add explanation through dissection of the simulation in
Sect.~\ref{sec:explanation}}.  The next section presents our
assumptions, input data, and methods.

%===========================================================================
\begin{table}
\caption{Line parameters.}
\label{tab:lines}
\centering
\begin{tabular}{lcc}
\hline \hline
Line                   & \MnI            & \FeI       \\
\hline
Wavelength [\AA]       & 5394.677        & 5395.215    \\            
Transition  &  $a^6S_{5/2}-z^8P^0_{7/2}$ & $z^5G^0_{2}-g^5F_{1}$ \\
Excitation energy [eV] & 0.0             & 4.446      \\
Oscillator strength ($\log \gf$)  & $-3.503$  & $-1.74$ \\
\rev{Land\'e factor}         & \rev{1.857}    & \rev{0.500} \\
$\cal{A}_{\rm lower}$             & $-2.41$   & $-$   \\
$\cal{A}_{\rm upper}$             & 18.23     & $-$  \\
Ionization energy [eV]            & 7.44      & 7.87 \\  %RR from my AQ
Abundance ($\log N_\rmH \is 12$)  & 5.35      & 7.51 \\  
\hline
\end{tabular}
\end{table}
%===========================================================================

%%%%%%%%%%%%%%%%%%%%%%%%%%%%%%%%%%%%%%%%%%%%%%%%%%%%%%%%%%%%%%%%%%%%%%%%%%%%
\section{Assumptions and methods}  \label{sec:method}
%%%%%%%%%%%%%%%%%%%%%%%%%%%%%%%%%%%%%%%%%%%%%%%%%%%%%%%%%%%%%%%%%%%%%%%%%%%%
%%%%%%%%%%%%%%%%%%%%%%%%%%%%%%%%%%%%%%%%%%%%%%%%%%%%%%%%%%%%%%%%%%%%%%%%%%%%
\subsection{Line selection}
%%%%%%%%%%%%%%%%%%%%%%%%%%%%%%%%%%%%%%%%%%%%%%%%%%%%%%%%%%%%%%%%%%%%%%%%%%%%
%% why these lines
We started this project with a wider line selection but for the sake
of clarity and conciseness we limit our analysis  to
\Mnline\ and the neighboring \Feline\ line, following the example of
  \citet{2007msfa.conf..189D} %C Danilovic++ magnetic source MnI
who show and modeled the activity modulation of both lines in
parallel.  The \Feline\ line serves here as prototype for all
comparable narrow lines.  The line parameters are given in
Table~\ref{tab:lines}.  The \MnI\ hyperfine structure constants ${\cal
A}$ come from
  \citet{2005ApJS..157..402B}. % Blackwell-Whitehead TEMP.BIB
All other values were taken from the NIST database at URL
\url{http://physics.nist.gov/asd3}.

%%%%%%%%%%%%%%%%%%%%%%%%%%%%%%%%%%%%%%%%%%%%%%%%%%%%%%%%%%%%%%%%%%%%%%%%%%%%
\subsection{Line synthesis}
%%%%%%%%%%%%%%%%%%%%%%%%%%%%%%%%%%%%%%%%%%%%%%%%%%%%%%%%%%%%%%%%%%%%%%%%%%%%
We perform line synthesis in the presence of magnetic fields 
with the code of
  \citet{2008ApJ...675..906S} %C Sanchez++: QS B measurement from HFS
which solves the radiative transfer equation for polarized light in a
given one-dimensional atmosphere via a predictor-corrector method.  It
yields the full Stokes vector, but we only use the intensity here.
The code includes evaluation of the Zeeman pattern for lines with
hyperfine structure using the routine of
  \citet{1978A&AS...33..157L}. % Landi 
This pattern depends on the magnetic field and on the hyperfine
structure constants, the quantum numbers of the upper and lower level,
the relative isotopic abundance, and the isotope shifts. The
splitting depends on the hyperfine structure constants $\cal{A}$ and
$\cal{B}$, which account for the two first terms of the Hamiltonian
describing the interaction between the electrons in an atomic level
and the nuclear magnetic moment.  $\cal{A}$ quantifies the
magnetic-dipole coupling, $\cal{B}$ the electric-quadrupole
coupling.  We consider $\cal{B}$ negligible here, again following
  \citet{2008ApJ...675..906S} %C Sanchez++: QS B measurement from HFS
who successfully reproduced multiple \MnI\ line profiles with
different HFS patterns.

%===========================================================================
%% Fig.\,\ref{fig:onedimmodels} 
%===========================================================================
\begin{figure}
  \centering \includegraphics[width=75mm]{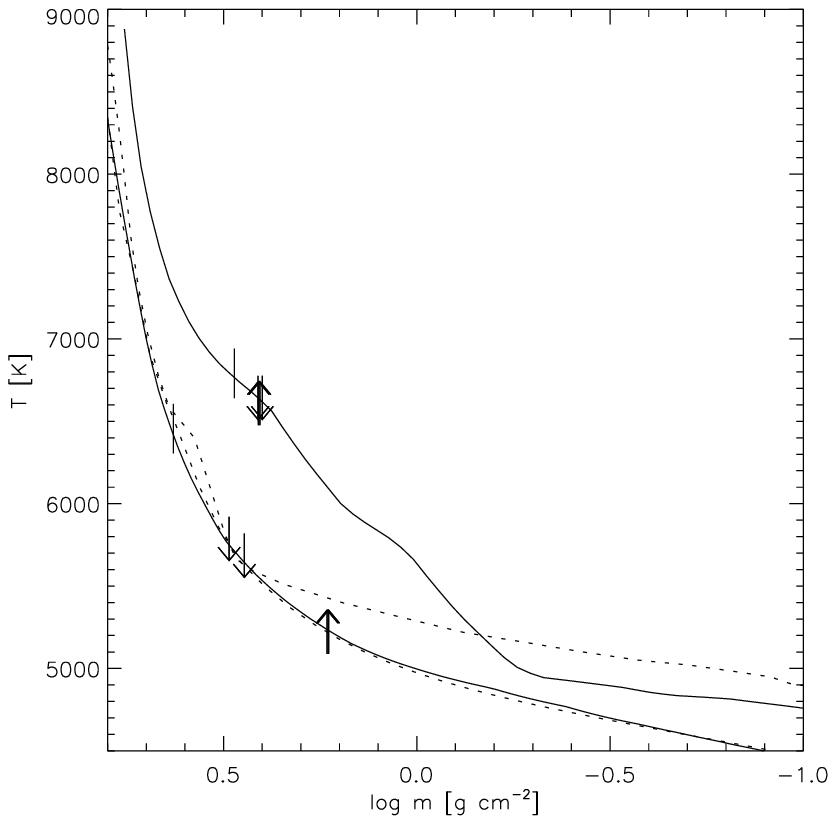}
  \caption[]{
  One-dimensional standard models 
\revpar
%% that are used or discussed in this paper, 
plotted as temperature against column mass. \revpar The
  \rev{lower solid} curve is the quiet-sun \rev{MACKKL} model of
%RR OOPS! was wrongly spelled all over the place
%
    \citet{1986ApJ...306..284M}; % MACKKL
  the \rev{upper solid} curve is the \rev{PLA model for fluxtubes in
  plage} of
    \citet{1992A&A...262L..29S}. %T Solanki+Brigljevic, fluxtubes
\revpar  
  The tick marks on these two curves specify $\tau_\lambda \is 1$
  locations for the line-center \rev{wavelength} of \Mnline\
  (\rev{upward arrows}), the line-center \rev{wavelength} of \Feline\
  (\rev{downward arrows}), and the continuum in between
  (\rev{arrowless}).  
%RR no nominal here since no Dopplershifts, only turbulent broadening
  Extra $\tau_\lambda \is 1$ ticks are added for \rev{line synthesis}
  without turbulent smearing, but \rev{these differ significantly only
  for the \FeI\ line and MACKKL}.  The two dotted models were defined
  by
  \citet{1999A&A...345..635U} %C Unruh++ irradiance modeling
  as basis for solar irradiance modeling \rev{and are discussed in
  Sect.~\ref{sec:discussion}}.  \revpar \rev{The lower one, close to
  MACKKL, is for quiet sun, the upper one for plage.}
}  
\label{fig:onedimmodels}
\end{figure}
%===========================================================================

%===========================================================================
%% Fig.\,\ref{fig:onedimdemo} 
%===========================================================================
\begin{figure}
  \centering \includegraphics[width=75mm]{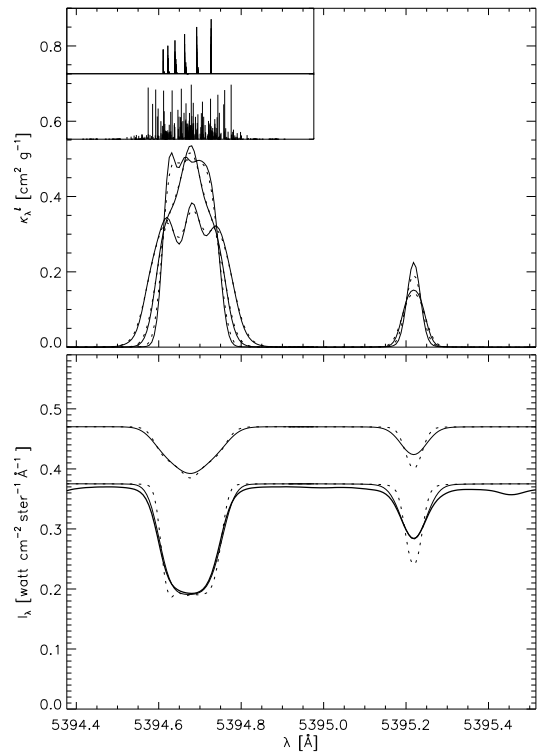}
  \caption[]{
  {\em Upper panel\/}: line extinction coefficient
  \rev{$\alpha_\lambda^\rml$} for \Mnline\ (left) and \Feline\ (right), for
  temperature $T \is 5000$~K and longitudinal magnetic field strength
  $B \is 0$~G (top solid curve), and 2000~G (lowest solid curve).  For
  the \MnI\ line there is also an intermediate profile for $B=1000$~G.
  The dotted curves result when temperature $T \is 7000$~K is inserted
  into the Dopplerwidth.  The insets specify the splitting pattern of
  \Mnline\ for $B \is 0$~G (upper) and $B \is 2000$~G (lower).
  {\em Lower panel\/}: line profiles resulting from classical
  one-dimensional spectral synthesis.  The lower solid curve in the
  lower trio is the quiet-sun disk-center spectrum in the atlas of
  \citet{Neckel1999} %T SP, FTS atlases "announcement", not in ADS
  which is an absolute-intensity version of the NSO/FTS atlas of
%
%%  \citet{2007assp.book.....W}. % Wallace+Hinkle+Livingston visatl
%RR bad latex - needs math for cm^-1; so old books.bib item
    \citet{Wallace+Hinkle+Livingston1998}. % visatl 
  The other solid curve in the lower trio is the spectrum computed
  from the MACKKL model including micro- and macroturbulence.  The
  dotted curve results when these fudge parameters are omitted.  The
  two upper spectra result from the PLA model with $B=1000$~G,
  also with and without turbulent smearing. }
\label{fig:onedimdemo}
\end{figure}
%===========================================================================

%%%%%%%%%%%%%%%%%%%%%%%%%%%%%%%%%%%%%%%%%%%%%%%%%%%%%%%%%%%%%%%%%%%%%%%%%%%%
\subsection{Assumption of LTE}  \label{sec:LTE}
%%%%%%%%%%%%%%%%%%%%%%%%%%%%%%%%%%%%%%%%%%%%%%%%%%%%%%%%%%%%%%%%%%%%%%%%%%%%

The source function of \Feline\ is likely to share the characteristic
properties of weak subordinate \FeI\ lines to possess LTE source
functions (with the equilibrium maintained by the much stronger
\FeI\ resonance lines) while having less-than-LTE opacity in the upper
photosphere in locations with steep temperature gradients (as controlled
by radiative overionization of iron in the near ultraviolet;
  see review by \cite{Rutten1988b}). %T Capri IAU Viotti review 

The source function of \Mnline\ should also remain rather close to LTE
because it is an intersystem transition with rather small 
oscillator strength, although not as forbidden as the
well-known LTE \MgI\ 4571.1~\AA\ line.
%RR what is log gf 4571?
Indeed, \Mnline\ closely obeys LTE in the detailed NLTE computations by
  \citet{2007A&A...473..291B} %C Bergemann+Gehren classical NLTE
for a standard solar atmosphere model assuming radiative equilibrium.
Its opacity is set by the ground-state population which is close to
LTE everywhere in these computations, even when photoionization is
increased by a factor 5000.  The degree of manganese ionization is
likely to exceed LTE in locations with steeper radial temperature
gradients, but so will the degree of of iron ionization; there is no
reason to suspect large difference between \MnI\ and \FeI\ line
formation with respect to NLTE effects.  Thus, apart from their large
hyperfine structure, \MnI\ lines should not behave very differently
from \FeI\ lines.  

The radial temperature gradients within magnetic concentrations are
probably close to local radiative equilibrium throughout their
photospheres
  (\cite{2005A&A...437.1069S}), %T Sheminova++ catube 
so that \rev{NLTE overionization is likely to affect 
both lines similarly also in these}.

%%%%%%%%%%%%%%%%%%%%%%%%%%%%%%%%%%%%%%%%%%%%%%%%%%%%%%%%%%%%%%%%%%%%%%%%%%%%
\subsection{One-dimensional modeling}  
%%%%%%%%%%%%%%%%%%%%%%%%%%%%%%%%%%%%%%%%%%%%%%%%%%%%%%%%%%%%%%%%%%%%%%%%%%%%
%% input models, micro and macro, B, HFS 

We use the standard model PLA for fluxtubes making up plage that was
derived in the 1980s by Solanki and coworkers from spectropolarimetry
of photospheric lines
 (\cite{1986A&A...168..311S}; %C Solanki, NET model (nr 1?)
  \cite{1988A&A...189..243S}; %C Solanki+Steenbock netw1,netw2,plage
  \cite{1992A&A...262L..29S}) %T Solanki+Brigljevic, fluxtubes
and shown in Fig.\,1 of 
  \citet{1993A&A...273..293B}. %C Bruls+Solanki
Since we only intend demonstration here, we do not apply spatial
averaging over upward-expanding and canopy-merging magnetostatic
fluxtubes as in
  \citet{1993A&A...268..736B}, %T Bunte+Solanki+Steiner, wine glass model
but simply use it as on-axis representation of a fully-resolved
fluxtube as in the cartoon in Fig.\,\ref{fig:BPcartoon}, with constant
field strength along the axis.  The same was done in Fig.\,4 of
  \citet{2008ApJ...675..906S}. %C Sanchez++: QS B measurement from HFS
PLA \revpar is shown in Fig.\,\ref{fig:onedimmodels}
together with the standard MACKKL quiet-sun model of 
  \citet{1986ApJ...306..284M} % MACKKL
which we use to represent the non-magnetic atmosphere outside
fluxtubes.  \rev{In a plot like this PLA appears to be much hotter than
MACKKL, but when PLA is shifted over the Wilson
depression in fluxtube modeling it is actually much cooler at equal
geometrical height.}  Figure~\ref{fig:onedimmodels} shows two additional
models from
  \citet{1999A&A...345..635U} %C Unruh++ irradiance modeling
that are discussed in Sect.~\ref{sec:discussion}.

%===========================================================================
%% Fig.\,\ref{fig:muram-images}
%===========================================================================
\begin{figure*}
  \sidecaption
  \includegraphics[width=120mm]{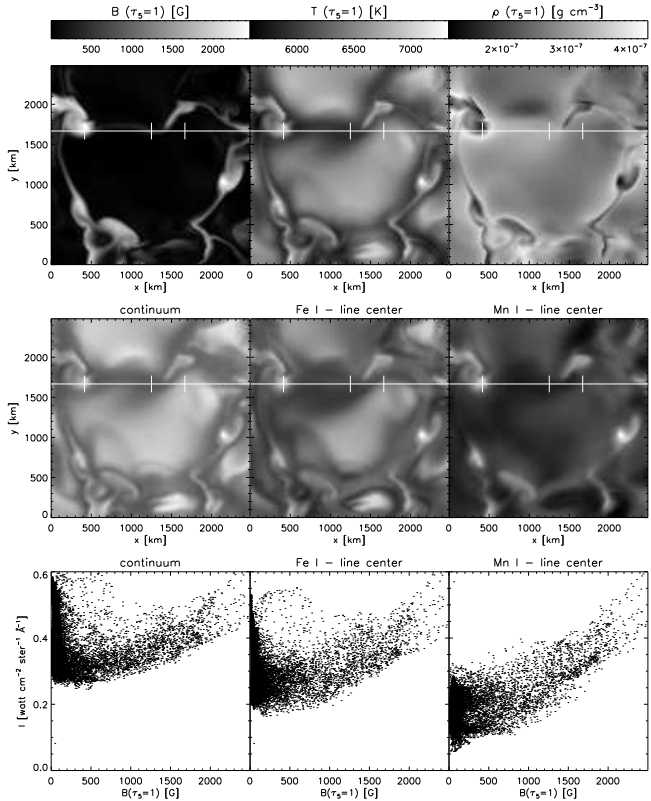}
  \caption[]{
  Results from the MURaM snapshot.
  {\em First row\/}: magnetic field strength, temperature, and gas
  density across the $\tau_5 \is 1$ continuum-escape surface.
  Greyscale calibration bars are shown on top.  The field is
  monopolar.  The white line specifies the cut used in
  Figs.~\ref{fig:muram-slices-A}--\ref{fig:muram-slices-C} to display
  formation parameters in a vertical plane through the simulation
  volume. It represents a spectrograph slit in
  Fig.\,\ref{fig:muram-profiles}.  The three superimposed ticks specify
  locations analyzed in detail in Fig.\,\ref{fig:muram-profiles}: a
  magnetic concentration, an intergranular lane, and the edge of the
  large granule in the center.
  {\em Second row\/}: synthetic intensity images for the continuum
  between the two lines and the \rev{nominal} \revpar wavelengths
  of \Feline\ and \Mnline.  They share the same intensity greyscale.
  These images represent predicted observations with a telescope
  matching the MURaM resolution by being diffraction-limited at 2.4-m
  aperture.
%RR 2 px MURaM px length into 1.22*lambda/D 
%RR DOT: print,1.22*500.*1E-9/0.45*180./3.14159*60.*60.
%RR MURaM: print,1.22*540E-9/(2.*(21./725.))*(180./3.14159)*60.*60.
%RR that is just the old LEST!
  The magnetic concentrations (point-like locations with the largest
  field strength) are about equally bright at all three wavelengths,
  but the non-magnetic granules (in the center and at the top of the
  field) are much darker in \Mnline\ than in the other two panels.
  {\em Third row\/}: the same intensities plotted as scatter diagrams
  per pixel against the magnetic field strength at the $\tau_5\is1$
  level for that pixel.  The upward tails of bright values at large
  field strength reach the same maxima in all three plots, but the
  granulation ($B < 200$~G) is much darker in \Mnline\ than in
  \Feline\ and the continuum.  The first two panels display a
  pronounced hook shape, but this hook is not present in the \MnI\
  panel.  Adding more field therefore gives more spatially-averaged
  brightening in the \MnI\ line than in the \FeI\ line and the
  continuum.  The \MnI\ line thanks its ``unhooking'' to its hyperfine
  structure through which it is less sensitive to the Doppler
  brightening that affects the \FeI\ line (and all similar lines) in
  the granulation.
}
\label{fig:muram-images}
\end{figure*}
%===========================================================================
%RR sidecaption format permits lots of prose in caption
%RN I had to cut the bb, so use my version

%%%%%%%%%%%%%%%%%%%%%%%%%%%%%%%%%%%%%%%%%%%%%%%%%%%%%%%%%%%%%%%%%%%%%%%%%%%%
\subsection{Three-dimensional simulation} 
%%%%%%%%%%%%%%%%%%%%%%%%%%%%%%%%%%%%%%%%%%%%%%%%%%%%%%%%%%%%%%%%%%%%%%%%%%%%
We use a single snapshot from a time-dependent simulation with the
MURaM (MPS/University of Chicago Radiative MHD) code
  (\cite{2003AN....324..399V}; % Voegler+Schuessler MURaM principles
   \cite{2004A&A...421..755V};  % Voegler  MURaM non-grey
   \cite{2005A&A...429..335V}). % Voegler++ details
\revpar
%% We give only a brief summary here.  
MURaM solves the three-dimensional
time-dependent MHD equations including non-local and non-grey
radiative transfer and accounting for partial ionization. 

The particular snapshot used here is taken from a simulation 
as the one described by
  \citet{2005A&A...429..335V}. % Voegler++
\revpar
Its horizontal extent is 6~Mm sampled in 288 grid points per axis, its
vertical extent 1.4~Mm with 100 grid points.  It was started with a
homogeneous vertical seed field of 200~G.
%% and then evolved during ?~minutes.  
Magnetoconvection gave it an appearance similar to active network 
with strong-field magnetic concentrations in intergranular lanes.  
A relatively quiet subcube was selected for \revpar our line synthesis.
It has $2.5 \times 2.5$~Mm horizontal extent and contains a large 
granule and a field-free intergranular lane, in addition to lanes 
containing magnetic concentrations of varying field strength 
(Fig.\,\ref{fig:muram-images}).

The line synthesis performed for this paper used the code described
above, treating the vertical columns in the subcube as independent
lines of sight.  \rev{The displays below are restricted to the nominal
NIST wavelengths of the two lines in Table~\ref{tab:lines},
corresponding to the line centers in a spatially-averaged disk-center
intensity atlas as in Fig.~\ref{fig:onedimdemo}}.

%%%%%%%%%%%%%%%%%%%%%%%%%%%%%%%%%%%%%%%%%%%%%%%%%%%%%%%%%%%%%%%%%%%%%%%%%%%%
\section{Results}  \label{sec:results}
%%%%%%%%%%%%%%%%%%%%%%%%%%%%%%%%%%%%%%%%%%%%%%%%%%%%%%%%%%%%%%%%%%%%%%%%%%%%

%%%%%%%%%%%%%%%%%%%%%%%%%%%%%%%%%%%%%%%%%%%%%%%%%%%%%%%%%%%%%%%%%%%%%%%%%%%%
\subsection{One-dimensional demonstration}  \label{sec:onedimdemo}
%%%%%%%%%%%%%%%%%%%%%%%%%%%%%%%%%%%%%%%%%%%%%%%%%%%%%%%%%%%%%%%%%%%%%%%%%%%%

Figure~\ref{fig:onedimdemo} presents results of our one-dimensional
modeling.  The solid curves in the upper panel show the spectral
variation of the extinction coefficient for both lines for temperature
$T = 5000$~K and field strengths $B=0$~G and $B=2000$~G, plus the
intermediate profile for $B=1000$ for the \MnI\ line.  For the \FeI\
line the Zeeman effect produces simple broadening but for the \MnI\
line the many hyperfine \revpar components, each with \rev{its} own
magnetic splitting, cause an intricate 
\revpar
%% multi-component 
pattern shown
in the lower part of the inset.  The resulting profile widens in the
wings but the core first becomes peaked at $B \is 1000$~G and then
splits into three collective peaks at $B \is 2000$~G.

The dotted curves result from inserting temperature $T=7000$~K into
the Dopplerwidth but not into other variables, in order to show the
effect of larger thermal broadening while keeping all other things
equal.  The \FeI\ profile for $B=0$~G looses appreciable amplitude
while the \MnI\ profile does not.

The lower panel of Fig.\,\ref{fig:onedimdemo} shows emergent intensity
profiles for the two lines.  The lowest solid curve is the observed
spatially-averaged disk-center spectrum.  It is closely matched by the
MACKKL modeling when applying \rev{standard} microturbulence (1~\kms)
and best-fit \revpar macroturbulence (1.28~\kms\ for \Mnline,
1.55~\kms\ for
\Feline).  When this artificial broadening is not applied the computed
\FeI\ line becomes too deep, but the depth of the
\MnI\ line does not change thanks to its flat-bottomed core.
The upper curves result from the PLA model with \revpar $B=1000$~G,
again with (solid) and without (dotted) turbulent smearing.  The
smearing again affects only the \FeI\ line.  \revpar It causes a
corresponding shift of the \FeI\ $\tau_\lambda \is 1$ location \rev{along
MACKKL} in Fig.\,\ref{fig:onedimmodels}.

Comparison of these MACKKL and PLA results shows that the spectrum
brightens at all wavelengths but most in the \MnI\ line, by a factor
2.  Turbulent smearing does not affect this line but it produces
large difference for the \FeI\ line.  In particular, if it is applied
to the MACKKL quiet-sun prediction but not to the PLA profile, the
\FeI\ line-center brightness increase is only a factor 1.4.
\revpar
%% much less than for the \MnI\ line.  

\rev{This difference in line-center brightening suggests} that
 \revpar the
\FeI\ line \rev{suffers more from} thermal broadening and the
thermodynamic fine structuring that was traditionally modeled with
micro- and macroturbulence. \rev{The apparent sensitivity of the \MnI\
line to magnetic activity may therefore indeed result from
non-magnetic quiet-sun Doppler smearing of the
\FeI\ line, lessening the latter's sensitivity.}

%===========================================================================
%% Fig.\,\ref{fig:muram-slices-A}
%===========================================================================
\begin{figure}
  \centering \includegraphics[width=88mm]{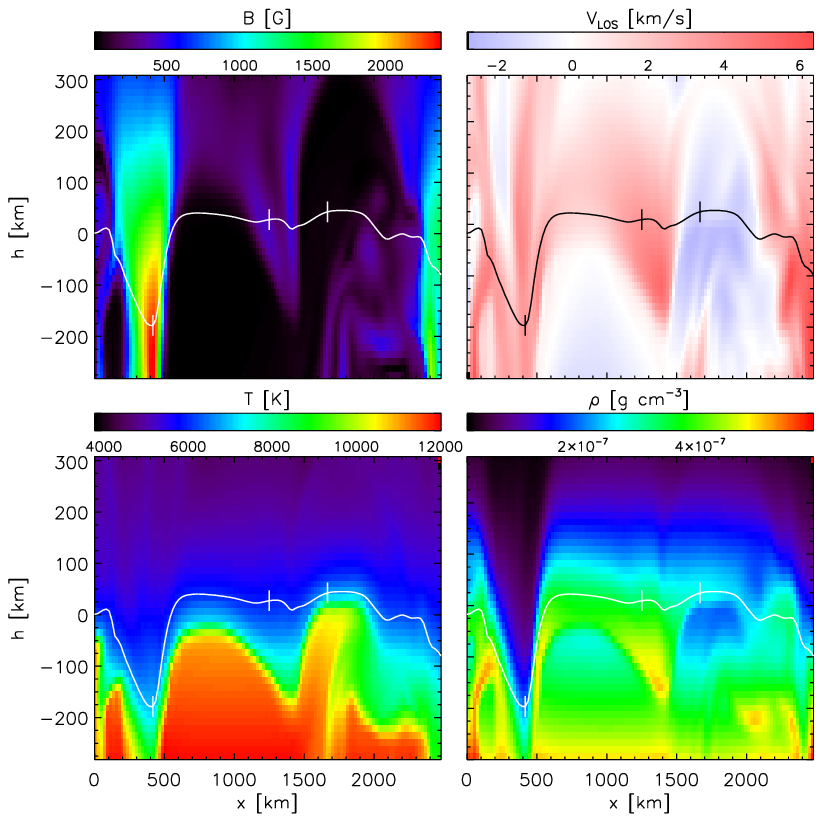}
  \caption[]{
  Various quantities in the MURaM snapshot across the vertical plane
  marked by the white horizontal line in Fig.\,\ref{fig:muram-images}.
  The upper panels show the magnetic field strength and radial
  velocity as a function of geometrical height versus horizontal
  location along the cut.  The height scale has $h=0$~km \rev{at the
  mean $\tau_5\is1$ level, averaged over all columns.  It extends
  882~km below and 518~km above that level in the simulation} but only
  the pertinent height range is shown here.  The velocity coloring is
  red for downdraft (redshift) and blue for updraft (blueshift).  The
  lower panels show the temperature and the gas density.  The overlaid
  curves specify $\tau_5\is1$ continuum formation depths.  The three
  ticks mark the locations analyzed in Fig.\,\ref{fig:muram-profiles}.
}
\label{fig:muram-slices-A}
\end{figure}
%===========================================================================
%RN I cut the bb, use my version 

%%%%%%%%%%%%%%%%%%%%%%%%%%%%%%%%%%%%%%%%%%%%%%%%%%%%%%%%%%%%%%%%%%%%%%%%%%%%
\subsection{Three-dimensional verification} \label{sec:simulation}
%%%%%%%%%%%%%%%%%%%%%%%%%%%%%%%%%%%%%%%%%%%%%%%%%%%%%%%%%%%%%%%%%%%%%%%%%%%%

The results from the MURaM simulation are shown in
Fig.\,\ref{fig:muram-images}. \revpar The three panels in the upper
row \revpar show basic state parameters across the $\tau_5\is1$
surface where $\tau_5$ is the continuum optical depth at $\lambda \is
5000$~\AA\ \rev{determined separately for each simulation column}.
The middle row displays synthetic intensity images for our three
diagnostics: the continuum between the two lines and the nominal
line-center wavelengths of
\Feline\ and \Mnline.  The bottom row shows these intensities in the
form of pixel-by-pixel scatter plots against the magnetic field
strength at $\tau_5\is1$.

The three images demonstrate directly why \Mnline\ shows larger
brightness contrast between non-magnetic and magnetic areas: \revpar
the magnetic bright points reach similar brightness in \rev{all} three
\rev{panels} but the granulation is markedly darker in this line.
Darker granulation implies larger sensitivity to \rev{activity, \ie\
addition of more magnetic bright points}.

The three scatter plots in the bottom row of
Fig.\,\ref{fig:muram-images} quantify this behavior.  In the continuum
\rev{plot} at left the darkest pixels lie in field-free or weak-field
intergranular lanes.  Addition of magnetic field within the lanes
brightens them, more for stronger fields, and for the strongest fields
almost up to the maximum brightness of field-free granular centers.

Note that in slanted near-limb viewing the magnetic concentrations do
not add brightness to dark intergranular lanes but permit deeper
viewing into bright granules behind them, adding brightness to the
already brightest features and so making faculae \rev{brighter} than the
granular background (see Fig.\,\ref{fig:BPcartoon}).

The scatter diagram for \Feline\ (bottom-center panel) shows a similar
hook shape.  The cloud of granulation pixels at left lies lower since
the line is an absorption line.  However, the pixels with the
strongest field still brighten to the same values as in the continuum
panel, implying that the line vanishes completely at \rev{its nominal}
wavelength.

The scatter diagram for \Mnline\ shows similar behavior, but it loses
the hook shape.  The line is yet darker in field-free granulation.  In
this case the corresponding dark cloud of \rev{low-field} pixels at
left does not have an upward tail.
%RR actually yet better for tau_lambda=1 B sampling but too complicated
However, the upward tail of pixels with increasing field still
stretches all the way from the dark lanes to the continuum values,
which again implies line vanishing.
  
\rev{If magnetic field is added to 
field-free granulation, this addition takes away points from the
bottom left of the distribution (corresponding to dark intergranular
lanes) and adds ``magnetic bright points'' at the upper right.  Its
effect is more dramatic in the ensemble average of the \MnI\ line than
in that of the \FeI\ line because the latter already has bright
contributions from field-free granules; in the \FeI\ line the
\rev{field-free granulation covers the same brightness range as the
field-filled lanes}.  Thus, field addition means larger}
spatially-averaged brightness increase in the \MnI\ line than in the
\FeI\ line.

We conclude that the MURaM synthesis duplicates the sun in having
larger brightness response to more activity in the \MnI\ line than in
the \FeI\ line -- not via chromospheric emission but through showing
granulation darker.

%%%%%%%%%%%%%%%%%%%%%%%%%%%%%%%%%%%%%%%%%%%%%%%%%%%%%%%%%%%%%%%%%%%%%%%%%%%  
\subsection{Explanation} \label{sec:explanation}
%%%%%%%%%%%%%%%%%%%%%%%%%%%%%%%%%%%%%%%%%%%%%%%%%%%%%%%%%%%%%%%%%%%%%%%%%%%

Unlike the sun, the MURaM simulation offers the opportunity to not
only inspect the emergent spectrum but \rev{also} to dissect the
behavior of pertinent physical parameters throughout the simulation
volume.  So instead of ending this paper here with the \revpar above
conclusion that neither a chromosphere nor NLTE coupling to \MgII\
\hk\ are needed to reproduce the activity sensitivity of \Mnline, we
\rev{add} analysis of the MURaM results \revpar to diagnose why the
granulation appears darker in the \MnI\ line -- or rather, why it
appears brighter in the \FeI\ line -- and so earn our use of the term
``explanation'' in the title of this paper.

\begin{figure*}
  \sidecaption
  \includegraphics[width=120mm]{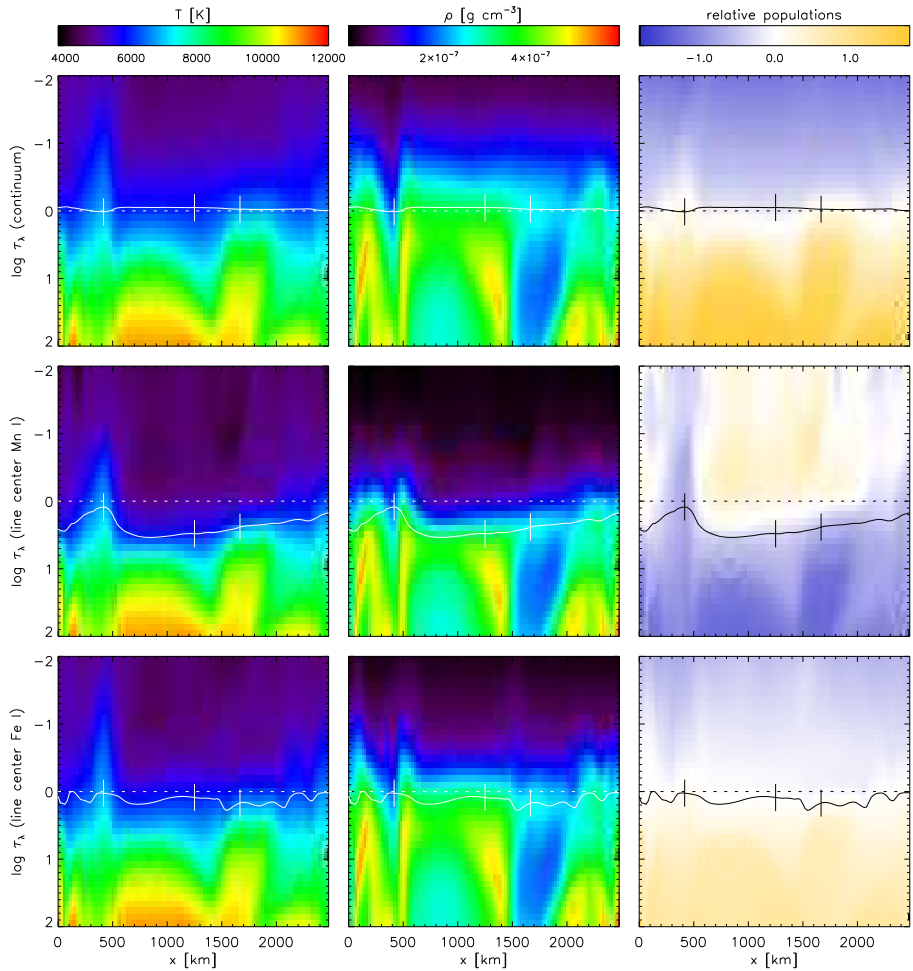}
  \caption[]{
  Temperature, gas density, and population variations across the
  vertical cut through the MURaM simulation plotted per column on
  radial optical depth scales for the nominal continuum (top row),
  \Mnline\ (middle row), and \Feline\ (bottom row) wavelengths.  The
  solid curves specify $\tau_5 \is 1$ depths, the horizontal dotted
  lines feature-specific $\tau_\lambda \is 1$ depths.  The third
  column displays fractional population offsets.  These are the
  lower-level populations for \Hmin\ bound-free transitions, the \MnI\
  line, and the \FeI\ line, each normalized by the total element
  density (hydrogen, manganese, iron) per location and shown in
  logarithmic units scaled to the mean value at $\tau_\lambda \is 1$
  along the cut.  Positive offsets are colored amber, negative bluish.
  The scale runs from $-1.8$ to $+1.8$.
}
\label{fig:muram-slices-B}
\end{figure*}

\begin{figure*}
  \sidecaption
  \includegraphics[width=120mm]{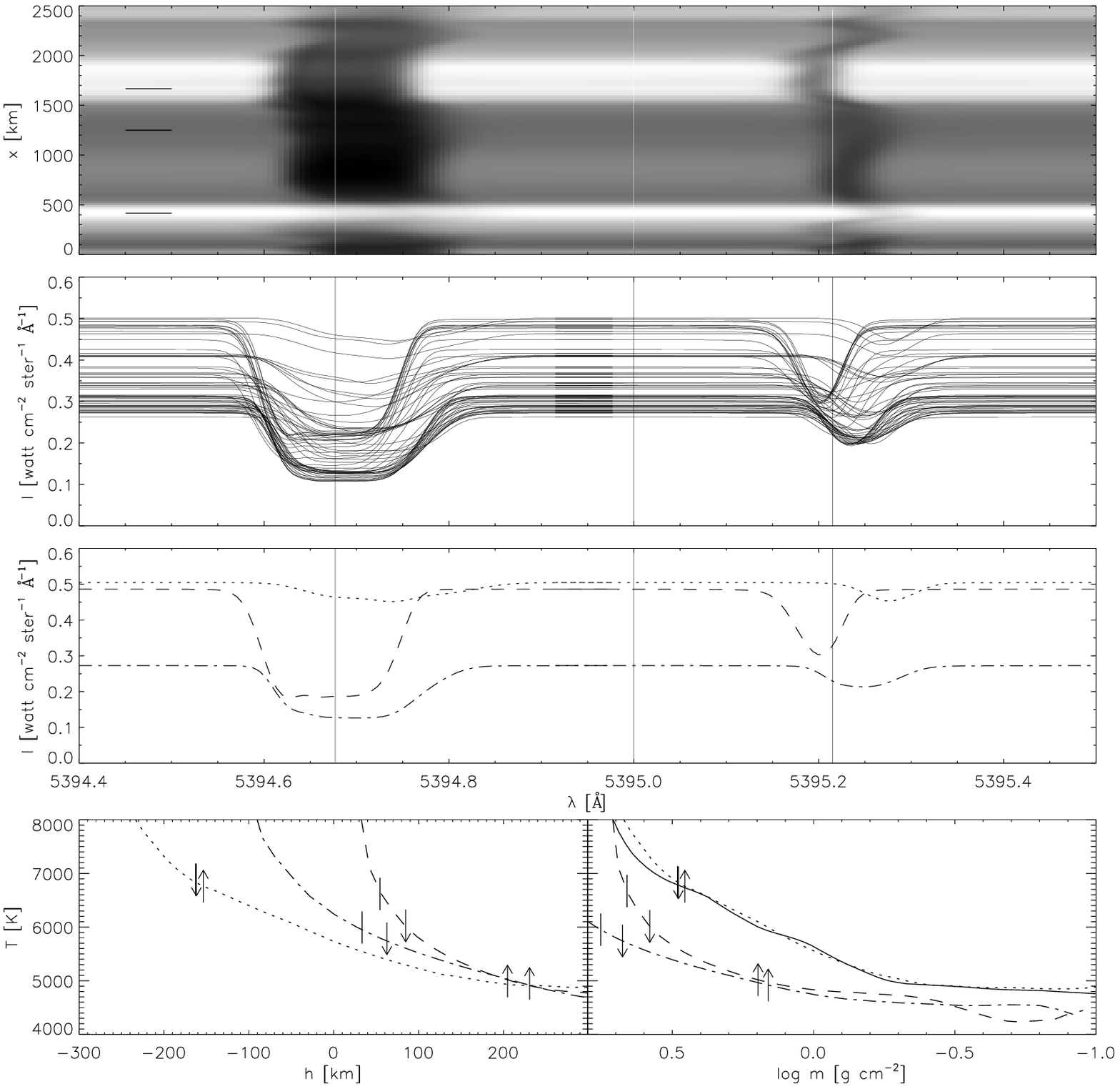}
  \caption[]{
  Further analysis of the formation of \Mnline, \Feline, and the
  intermediate continuum in the MURaM simulation.
  {\em Top panel\/}: spectrum synthesized from the computed emergent
  intensities.  The vertical coordinate corresponds to the $x$
  sampling along the cut specified \revpar in
  Fig.\,\ref{fig:muram-images}.  The three black horizontal markers
  correspond to the ticks selecting the bright point (lowest),
  field-free intergranular lane (middle), and granule edge (highest).
  The three \revpar vertical lines specify the nominal
  wavelengths for \Mnline\ (left), \Feline\ (right), and the
  intermediate continuum (center).
  {\em Second panel\/}: spectral profiles for every second pixel
  along the cut.
  {\em Third panel\/}: spectral profiles for the three locations
  specified by ticks in Fig.\,\ref{fig:muram-images} and black markers
  in the top panel.  {\em Dotted\/}: magnetic concentration.  {\em
  Dot-dashed\/}: intergranular lane.  {\em Dashed\/}: granule edge.
  {\em Bottom panels\/}: temperature stratifications for the three
  selected locations, \rev{at left against geometrical height, at
  right against column mass as in Fig.~\ref{fig:onedimmodels}.  The
  solid curve is the PLA model, as in Fig.~\ref{fig:onedimmodels}.
  The tick marks specify the $\tau_\lambda \is 1$ locations for the
  nominal \rev{wavelengths} of \Mnline\ (\rev{upward arrows}), of
  \Feline\ (\rev{downward arrows}), and of the continuum in between
  (\rev{arrowless})}.  \rev{In the magnetic concentration the \Feline\
  tick coincides nearly with the continuum one.  The Wilson depression
  is much larger for the \MnI\ line than for the continuum and the
  \FeI\ line.}
\revpar
}  
\label{fig:muram-profiles}
\end{figure*}

\begin{figure*}
  \sidecaption
  \includegraphics[width=120mm]{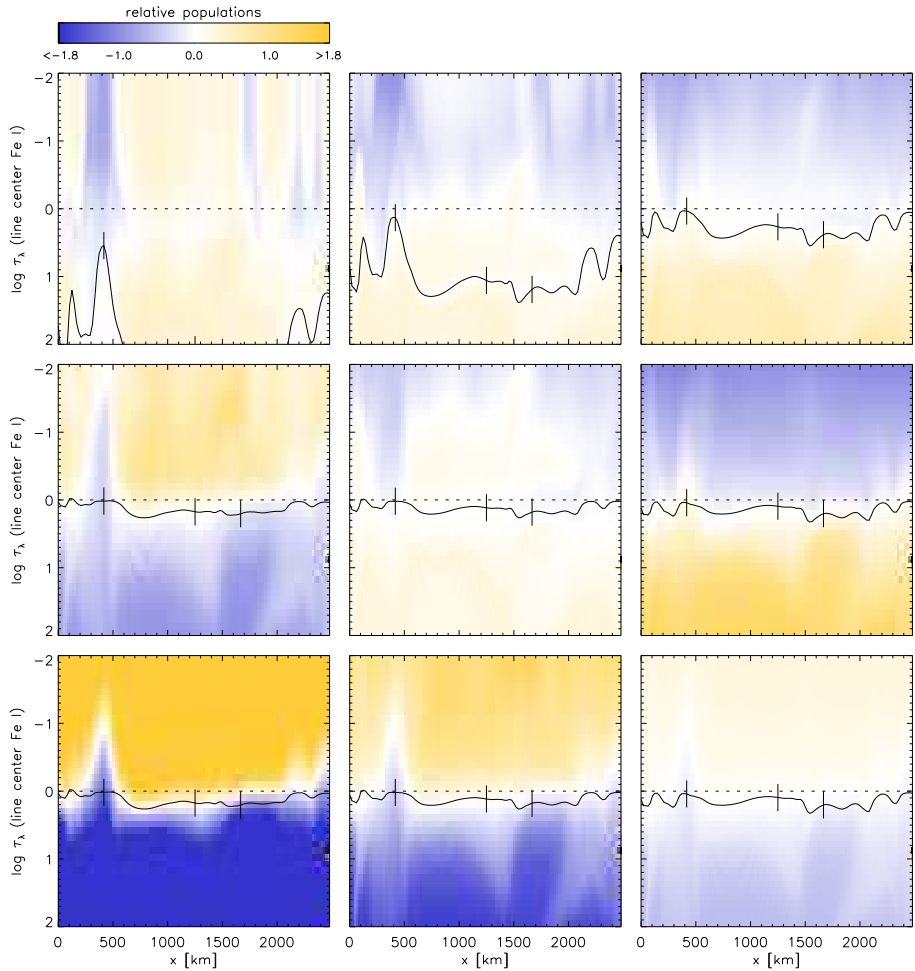}
  \caption[]{
  The effect of excitation energy and line strength on \FeI\ line
  formation.  Each panel is the fractional population of the lower
  level of an artificial line comparable to \Feline\ along the cut
  through the MURaM simulation defined in
  Fig.\,\ref{fig:muram-images}, again shown on \rev{nominal}
  line-center optical depth scales, normalized to the mean value along
  the dotted line, and with the same logarithmic color coding as in
  the third column of Fig.\,\ref{fig:muram-slices-B}, \rev{ clipped at
  offsets $-1.8$ and $+1.8$.}  The solid curves are again the $\tau_5
  \is 1$ locations for the continuum at 5000~\AA, the dashed lines the
  line-center $\tau_\lambda \is 1$ formation indication.
  {\em Top row\/}: the \Feline\ line at excitation energy 2, 3, and 4~eV,
  respectively.  The corresponding mean Boltzmann factors at
  $\tau_\lambda \is 1$ are $1.2\times10^{-3}$, $1.7\times10^{-4}$, 
  and $5.2\times10^{-5}$.
  {\em Second row\/}: the \Feline\ at excitation energy 0, 3, and 6~eV
  but with the oscillator strength scaled to maintain the same line
  strength in the emergent spectrum that \Feline\ shows (by
  $25\times10^{-3}$, 25 and $25\times10^{3}$, respectively).
  {\em Bottom row\/}: idem, but the color coding now measures
  deviations from the Milne-Eddington approximation by showing the
  lower-level population divided by the \Hmin\ population.  
  The leftmost panel is severely clipped.  The Milne-Eddington
  approximation improves dramatically with higher excitaton.  
  \revpar

}
\label{fig:muram-slices-C}
\end{figure*}

The white line with three ticks in the grey-scale panels of
Fig.\,\ref{fig:muram-images} specifies the spatial samples that are
selected in Figs.~\ref{fig:muram-slices-A}--\ref{fig:muram-slices-C}.
This particular cut and the tick locations were chosen to sample a
strong-field magnetic concentration that appears as a bright point in the
intensity images (lefthand tick), a non-magnetic dark lane (middle
tick), and a granule (righthand tick).  Unfortunately, this cut samples
only the edge rather than the center of the large granule covering the
center part of the field but otherwise we wouldn't have sampled both
a bright point and a non-magnetic lane.  \revpar A magnetic lane
less extreme than the bright point is sampled at the far right.

%RR adding curves for that might be nice but would clutter Fig 7 too much 

Figure~\ref{fig:muram-slices-A} diagnoses MURaM physics in the
vertical plane defined by this cut.  It shows behavior that is
characteristic of solar magnetoconvection near \revpar the surface.
The overlaid curves specify the $\tau_5 \is 1$ \revpar continuum
surface.  The very low gas density (4th panel) in the magnetic
concentration at $x \approx 400$~km produces a large Wilson depression
of about 200~km.  The deep dip in the $\tau_5\is1$ curve so samples
relatively high temperature, as well as a relatively flat vertical
temperature gradient.  The intergranular lane at $x \approx 1300$~km
combines low temperature with high density and strong downdraft; the
granule edge at $x \approx 1650$~km combines higher temperature with
gentler updraft and low subsurface density.
%RR is that normal?  also in the center of the big granule?
%RR remains not-so-good sample but I see no better

Figure~\ref{fig:muram-slices-B} repeats this vertical-plane display of
temperature and density but plotted per column on the radial optical
depth scale that belongs to each diagnostic.  The panels in the first
and second columns show the atmosphere ``as seen'' by each spectral
feature \rev{at its nominal wavelength}.  The dotted horizontal lines
at $\log \tau_\lambda \is 0$ indicate their formation heights.  The
solid curves are \revpar the \revpar $\tau_5 \is 1$ locations.
\revpar The third column shows relative behavior of the
corresponding opacities in the form of fractional lower-level
population variations.

%===========================================================================
%% Fig.\,\ref{fig:muram-slices-B}
%===========================================================================
%===========================================================================

%===========================================================================
%% Fig.\,\ref{fig:muram-profiles}
%===========================================================================
%===========================================================================

The top panels of Fig.\,\ref{fig:muram-slices-B} show only slight
differences between the \revpar $\tau\is1$ locations along the cut.
%RR balance changes between Paschen bf and H-minus bf?
The magnetic concentration has an appreciable hump in
$T(\tau_\lambda)$ and \revpar dip in $\rho(\tau_\lambda)$ around
$\tau_5 \is 1$.  The relatively high temperature and low density there
combine into increased electron-donor ionization.  This is illustrated
by the sixth panel \revpar showing the fractional population variation
of the lower level of \Mnline, which is the \MnI\ ground state.  Its
behavior equals the depletion by manganese ionization apart from a
minor correction for the temperature sensitivity of the \MnI\
partition function.  Manganese has too small abundance to be an
important electron donor, but it ionizes similarly to iron
(Table~\ref{tab:lines}) so that this panel illustrates characteristic
electron-donor ionization, with neutral-stage depletion occurring
within magnetic concentrations and at large depth.  The corresponding
increase of the free-electron density produces larger \Hmin\ opacity,
evident \revpar as overall color gradient reversal between the top and
center panels in the third column.  \revpar It results in an upward
enhancement peak at the magnetic concentration in the \Hmin\
population panel.  This relative increase of the continuum opacity
explains that $\tau_\lambda \is 1$ is reached at lower density in the
magnetic concentration than in the adjacent intergranular lane (second
panel); the Wilson depression is smaller than one would estimate from
pressure balancing alone.  The flat temperature gradient in the
magnetic concentration produces a marked upward extension in
$T(\tau_\lambda)$ in the first panel.  It contributes brightness
enhancement along much of the intensity contribution function and so
makes magnetic concentrations appear bright with respect to
non-magnetic lanes.  However, the hook pattern in the continuum
scatter plot in Fig.\,\ref{fig:muram-images} shows that this lane
brightening does not exceed granular brightness.

The center row of Fig.\,\ref{fig:muram-slices-B} shows that \Mnline\ is
generally formed higher than the continuum, but not in the magnetic
concentration where the upward hump in the $\tau_5 \is 1$ curve nearly
reaches the $\tau_\lambda \is 1$ level.  This is because the
relatively high temperature and low density there increase the degree
of manganese ionization, as evident in the third panel which shows a
marked upward blue extension.  The line weakens so much that the
brightest \MnI\ pixels in Fig.\,\ref{fig:muram-images} reach the same
intensity as in \revpar the continuum, also sampling the upward
high-temperature extension (first column).
%RR plus a bit Dopplershift, see muram-profiles
Conversely, the largest line-to-continuous opacity ratio (\ie\ the
largest separation between the $\tau_\lambda \is 1$ and $\tau_5 \is 1$
locations) is reached along the intergranular lane where the
neutral-stage population is larger due to relatively large density and
low temperature.  Thus, the curve separation maps the variation
of the fractional ionization along the dotted line.  Since the
temperature increases inward anywhere, the deeper sampling within the
magnetic concentration and the higher sampling in the granulation,
especially in the lanes, together increase the brightness contrast
between bright point and granulation compared to that in the
continuum.  This effect of increased ionization is similar to the
effect of CH dissociation in the G band and reduced damping in the
\Halpha\ wings within magnetic concentrations.  It enhances the
magnetic lane brightening so that that exceeds granular brightnesses,
undoing the hook shape of the continuum scatter plot.

The bottom row of Fig.\,\ref{fig:muram-slices-B} shows the
corresponding plots for \Feline.  Iron and manganese ionize similarly
so that low-excitation \FeI\ lines suffer the same depletion in
magnetic concentrations.  However, \Feline\ is a high-excitation line;
its Boltzmann sensitivity to higher temperature largely compensates
for the enhanced ionization so that the panel in the third column
looks much more like the top one for \Hmin\ than like the second one
for \Mnline: the overall color gradient flips back again.  The line
again vanishes in the magnetic concentration, so that the
strongest-field pixels in Fig.\,\ref{fig:muram-images} again reach the
continuum intensities.  One might expect that the curve separation in the
bottom panels would follow the temperature pattern under the dotted
line in the first column, but this is not the case; for example, the
line also nearly vanishes \rev{(at its nominal wavelength)} in the
intergranular lane.  This discordant variation is caused by the
Dopplershifts imposed on the line extinction by the flows displayed in
the upper-right panel of Fig.\,\ref{fig:muram-slices-A}.  Comparison
shows that the unsigned amplitude of the flow variation along the cut
at $100-200$~km above the $\tau_5 \is 1$ curve there is mapped
precisely into reversed modulation of the $\tau_5 \is 1$ curve for
\Feline\ here.  \rev{Thus, the line vanishes because it is shifted aside
from its nominal wavelength.} 

We demonstrate the Doppler-related formation differences between the
two lines further in Fig.\,\ref{fig:muram-profiles}.  The upper two
panels display spectral representations along the cut defined in
Fig.\,\ref{fig:muram-images}, in the form of a spectral image and of
profile samples.  The top panel shows what would have appeared in an
observational spectrogram from a telescope with MURaM resolution.
\rev{Both lines show a bright ``line gap'' in the magnetic
concentration near the bottom.}  At $x \approx 2500$~km \revpar the
intergranular lane with less strong field still causes a noticeable
\revpar gap.  \rev{Large Dopplershifts occur in the field-free lane 
and the granular edge.}

The second panel shows the product of data reduction \rev{of this
spectrogram}.  Both panels illustrate that the line-center intensity
of the \MnI\ line is not very sensitive to Dopplershifts, whereas the
\FeI\ line shifts well away from its nominal \revpar wavelength nearly
everywhere.  It also weakens more in the granule edge from larger
thermal broadening (Fig.\,\ref{fig:onedimdemo}).  Taking the
spatial mean at each \rev{nominal} wavelength does a fair job of
intensity averaging for the \MnI\ line but misses nearly all dark
cores in the \FeI\ line, especially in the lanes but also in granules.
Note that this particular cut does not represent hot granules \revpar
well; more samples of these would add many profiles with weakened
cores blueward of the nominal \revpar \FeI\ wavelength.

The third panel shows the spectral profiles for \revpar
the three sample locations along the cut.  The \rev{nominal} \FeI\
line-center wavelength misses all three cores!  Thus, the brightness
average at this wavelength is much \rev{higher} than it would be for
an undisturbed line of this opacity.  The \MnI\ line-center
wavelength, however, only misses the deepest part of the
magnetic-concentration profile which is weak anyhow.  This disparity
in Doppler sensitivity explains why the granulation in
Fig.\,\ref{fig:muram-images} is much darker in \Mnline\ than in
\Feline.

\rev{Also} note the \revpar sharpening of the \MnI\ line profile
from boxy to more pointed in some of the intermediate profiles in the
second panel, which follows the extinction coefficient behavior in
Fig.\,\ref{fig:onedimdemo}.  \rev{It contributes to the
decrease in equivalent width of \Mnline\ at larger activity, which is
not analyzed here.}

The bottom panels of Fig.\,\ref{fig:muram-profiles} show the vertical
temperature stratifications at the three sample locations, at left
against geometrical height \rev{with $h \is 0$ at $\tau_5 \is 1$ for
the simulation mean}, at right against column mass per feature.  These
graphs \rev{link the simulation results back to the classical fluxtube
modeling in Sect.~\ref{sec:onedimdemo}}.  They \rev{display} familiar
properties of granulation and magnetic concentrations: the temperature
gradient is steepest in granules, flatter in intergranular lanes, with
a mid-photosphere cross-over producing ``reversed granulation'', and
flattest in fluxtubes.  The heights of formation, marked by ticks
specifying the $\tau_\lambda \is 1$ locations, are similar \rev{per
line} in the \rev{granular edge} and the lane but much deeper \revpar
in the magnetic concentration \rev{which is the coolest feature at
equal geometrical height but the hottest at equal column mass and at
the $\tau_{\rm cont} \is 1$ location per feature.  \rev{The Wilson
depression is about 200~km in the continuum and twice as large for the
\MnI\ line through enhanced ionization} within the magnetic concentration,
but not for the \FeI\ line due to its Doppler-deepened granulation
sampling.  Both lines weaken severely in the magnetic concentration
through this deep sampling, and both are redshifted away from their
nominal wavelength, \Feline\ all the way so that its $\tau_\lambda \is
1$ tick coincides with the continuum one.}
\revpar

%===========================================================================
%% Fig.\,\ref{fig:muram-slices-C}
%===========================================================================
%===========================================================================

%RN I still would like to diagnose the reason for the \lg and \ln
%RN difference between our equal-strength estimates (me kappa_l, you W).

%%%%%%%%%%%%%%%%%%%%%%%%%%%%%%%%%%%%%%%%%%%%%%%%%%%%%%%%%%%%%%%%%%%%%%%%%%%%
\section{Discussion}   \label{sec:discussion}
%%%%%%%%%%%%%%%%%%%%%%%%%%%%%%%%%%%%%%%%%%%%%%%%%%%%%%%%%%%%%%%%%%%%%%%%%%%%

\subsection{Comparison of 1D modeling and 3D simulation}
%%%%%%%%%%%%%%%%%%%%%%%%%%%%%%%%%%%%%%%%%%%%%%%%%%%%%%%%
\revpar
We have used both classical \rev{empirical} fluxtube modeling
(Sect.~\ref{sec:onedimdemo}) and modern MHD simulation
(Sect.~\ref{sec:simulation}).  \rev{The bottom panels of
Fig.\,\ref{fig:muram-profiles} connect the latter with the former.
The righthand panel shows remarkably close agreement between the PLA
model and the MURaM magnetic-concentration stratification.  This may
be taken as mutual vindication of these very different techniques, but
not as proof of correctness since both assumed LTE ionization
(discussed below) while the sun does not.}

The simulation resolves the \rev{granular} Dopplershifts that were
\revpar emulated by turbulence in \rev{the classical} modeling.
\revpar
\rev{The simulation analysis in Sect.~\ref{sec:explanation}}
\revpar confirms that \Mnline\ activity-brightens \rev{more than
\Feline\ } because its intrinsic hyperfine-structure broadening
 suppresses thermal and Doppler brightening
\rev{in non-magnetic granulation.
This was already indicated by the 1D demonstration in
Fig.~\ref{fig:onedimdemo}.}

\subsection{Validity of LTE}
%%%%%%%%%%%%%%%%%%%%%%%%%%%%
We have assumed LTE throughout this paper.  In Sect.~\ref{sec:LTE}
we noted that the most likely departure from LTE which may affect our
two lines is loss of opacity through radiative over\-ionization, in
locations with steep temperature gradients where the angle-averaged
intensity exceeds the Planck function in ultraviolet photo-ionization
edges
  (see \eg\ \cite{Rutten2003:RTSA} for more explanation). %T edition 2003 
The steepest temperature gradients in Fig.\,\ref{fig:muram-slices-A}
occur not along radial columns but transverse through the magnetic
concentration.  Therefore, the ionization which contributes to the
vanishing of the \MnI\ line from the corresponding bright point may be
underestimated.  If so, the \Hmin\ opacity is equally underestimated.
\rev{However}, the MURaM simulation itself assumes LTE with \revpar coarse
spectral sampling and \rev{is likely to differ intrinsically if NLTE
overionization affecting the \Hmin\ opacity is taken into account.
Such overionization was similarly neglected in the empirical
construction of the PLA model based on \FeI-line spectropolarimetry
assuming LTE to define the optical depth scales.  Improvement}
requires non-trivial evaluation of the ultraviolet line haze.

\subsection{Extension to other lines}
%%%%%%%%%%%%%%%%%%%%%%%%%%%%%%%%%%%%%
\revpar
We have \rev{demonstrated} and analyzed Doppler brightening in detail
only for \Mnline\ and \Feline. The first may be taken to exemplify all
lines with wide boxy extinction profiles, the second all lines with
narrow peaked profiles, \ie\ almost all other photospheric lines.
Figure~\ref{fig:muram-slices-C} \rev{serves to demonstrate the latter
generalization} by extending the MURaM synthesis to a range of
artificial \FeI\ lines by simply placing
\Feline\ at other excitation energies.
\revpar
%% both with and without the corresponding Boltzmann opacity increase.  
The first two rows again display relative population variations across
the simulation cut using the same color coding as in the third column
of Fig.\,\ref{fig:muram-slices-B}.  The top row of
Fig.\,\ref{fig:muram-slices-C} combines the effects of the Boltzmann
factor on the amplitude of the line extinction and on its temperature
sensitivity.  The second row isolates the latter by making each line
just as weak as \Feline.  

The first panel of Fig.~\ref{fig:muram-slices-C} is included only for
illustration because this artificial line is already too strong for
reliable synthesis from our simulation and would also suffer
appreciable departures from LTE if it existed in reality
  (see \cite{1982A&A...115..104R}). %C Rutten+Kostik
It is formed near the top of the simulation volume \rev{which
sets the columnar structure in the upper part;}
%RR ms had Doppler, was wrong, but don't know what sets it
the sizable vertical flows there impose the separation between the
two $\tau\is1$ curves. 
%RR this seems correct
In reality, very strong \FeI\ lines suffer
less convective and thermal Doppler weakening in their cores which
explains the observation by
  \citet{2004IAUS..223..645M} %C Malanushenko++ imaging in MnI
that the strongest \FeI\ lines show some (but only slight) global
activity modulation.  Even that modulation has nothing to do with the
chromosphere!  The second and third panel describe deeper line
formation with steeper gradients. Their $\tau\is1$ separations also
demonstrate Doppler modulation.  Thus, Doppler weakening 
affects these three lines just as it affects \Feline.

The second row of Fig.\,\ref{fig:muram-slices-C} illustrates the effect
of Boltzmann temperature sensitivity at given emergent line strength.
The first panel describes the \FeI\ ground state and is nearly the
same as the sixth panel of Fig.\,\ref{fig:muram-slices-B}.  Towards
higher excitation energy the Boltzmann increase compensates more of
the depletion by ionization.  In the center panel the population
offset gradients are very flat; at right, they are reversed so that
the population variation mimics the behavior of \Hmin\ in the third
panel of Fig.\,\ref{fig:muram-slices-B} closely, although at slightly
smaller amplitude.  The Doppler modulation of the $\tau \is 1$
separation remains \rev{similar}, irrespective of this gradient
reversal in fractional population distribution.  \rev{Thus, Doppler
brightening affects all photospheric \FeI\ lines similarly.}

The third row of Fig.\,\ref{fig:muram-slices-C} repeats this test in
terms of the Milne-Eddington approximation (constant line-to-continuum
opacity ratio with height) which is often assumed in photospheric
polarimetry
  (\eg\ 
  \cite{1977SoPh...55...47A}; % Auer++ ME Stokes fitting
  \cite{1987ApJ...322..473S}; % Skumanich+Lites better ME
  \cite{1998ApJ...494..453W}; % Westendorp++ compare ME-SIR
  \cite{2007ApJ...670L..61O}). % Oroczo++ Hinode/SOT ME 
%
%RR FeI 6303 are at 3.7 eV, not yet very good
At left the combination of \FeI\ depletion by ionization and the
corresponding increase in \Hmin\ opacity implies squaring of the
population offsets or twice steeper gradients in this logarithmic plot
than in the second row.  The Milne-Eddington approximation fails
badly but least within the magnetic concentration.  It improves
with excitation energy through the Boltzmann compensation.  The panel
at right shows the residue from subtracting the sixth panel in this
figure from the third panel in Fig.\,\ref{fig:muram-slices-B},
resulting in rather small deviations.  
\revpar
This behavior was shown earlier in Fig.\,3 of
  Rutten \& van der Zalm (1984; 
  \nocite{1984A&AS...55..143R} % RR+vanderZalm 
reprinted in Fig.\,9.2 of
  \cite{Rutten2003:RTSA}). %T edition 2003 

\subsection{Irradiance modeling}
%%%%%%%%%%%%%%%%%%%%%%%%%%%%%%%%
In this paper we have not made the step from radial viewing to
full-disk averaging.  The demonstration in Sect.~\ref{sec:onedimdemo}
used Solanki's one-dimensional empirical model for a fluxtube in plage
but without adding flaring fluxtube geometry, granular presence next
to a fluxtube and behind it in slanted facular viewing, multiple-tube
interface geometry, spatial averaging over these geometries, the
multi-angle evaluation needed to emulate center-to-limb viewing, and
without full-disk averaging and consideration of spatial distributions
over the solar surface and their variations with the solar cycle.  All
these non-trivial aspects would need careful quantification \revpar to
\rev{expand} the demonstration of enhanced sensitivity \rev{in
Sect.~\ref{sec:onedimdemo}} into a quantitative estimate for
comparison with Livingston's irradiance data.  The simulations in
Sect.~\ref{sec:simulation} yielded profile synthesis for a small area
of solar surface containing strong-field concentrations that may be
considered more realistic than idealized magnetostatic fluxtubes but,
nevertheless, 
\revpar
%% the step from simulation to 
generation of full-disk signals comparable to \rev{Livingston's data}
%% the monitoring observations 
still requires \rev{all} the above quantifications.  This
effort is not done here.

In contrast, \rev{the} step from one-dimensional line synthesis to
emulation of the full-disk and cycle-dependent integrated signal was
recently made by
  \citet{2007msfa.conf..189D} %C Danilovic++ magnetic source MnI
using the SATIRE (Spectral And Total Irradiance Reconstruction)
approach of
  \citet{2000A&A...353..380F}, %C Fligge++ only cont
  \citet{2003A&A...399L...1K} %C Krivova++
and
  Wenzler et al. (2005, 2006).
    \nocite{2005A&A...432.1057W} %C Wenzler++
    \nocite{2006A&A...460..583W} %C Wenzler++
In this technique the spatial distributions of spots and plage are
extracted from full-disk magnetograms and continuum images to derive
disk-coverage distributions throughout the more recent activity
cycles.  Each component \rev{(quiet sun, plage, spots)} is represented by a
standard one-dimensional model atmosphere.  \rev{The first two 
are shown as dotted curves in Fig.~\ref{fig:onedimmodels}.
The lower one for quiet sun is the \revpar
radiative equilibrium model of
  Kurucz (1979, 1992a, 1992b)
    \nocite{1979ApJS...40....1K} % Kurucz models
    \nocite{1992RMxAA..23..181K} % Kurucz solar update
    \nocite{1992RMxAA..23..187K} % Kurucz solar update
which is nearly identical to \revpar MACKKL.
The upper one for plage \revpar was made by
  \citet{1999A&A...345..635U} %C Unruh++ irradiance modeling
by smoothing model P of
    \citet{1993ApJ...406..319F} % FAL including FALC model
\revpar and deleting its chromospheric temperature rise.  For spots a
Kurucz radiative-equilibrium model with low effective temperature is
used, not shown here.}
  \citet{2007msfa.conf..189D} %C Danilovic++ magnetic source MnI
found that using these models with the empirically established SATIRE
coverage fractions 
\revpar
%% (which reproduce the solar continuous irradiance
%% variations pretty well) 
gives a good reproduction of Livingston's data
both for \Mnline\ and \Feline.

\revpar \rev{Such} use of the dotted models in
Fig.~\ref{fig:onedimmodels} is \revpar an ad-hoc trick to reproduce
the larger brightness of plage.  The \Mnline\ line activity-brightens
more than \Feline\ in this simplistic modeling because it is a
stronger line, hence formed higher, hence getting more out of the
divergence between the two models with height.  Any line as strong
would show the same brightening.  Stronger lines would brighten
\revpar 
more; in particular, \FeI\ lines with deeper cores than \Mnline\ would
show larger activity modulation in conflict with the observations.

Actually, as illustrated in Fig.\,\ref{fig:BPcartoon} and demonstrated
in \mbox{Figs.~\ref{fig:muram-images}--\ref{fig:muram-slices-C}}, plage
thanks its \rev{disk-center} brightening in any photospheric
diagnostic not to being hotter at equal height but to
below-the-surface viewing of hot-wall heat within magnetic
concentrations.  The SATIRE modeling does not evaluate Wilson
depressions, but as long as the SATIRE models are used in
one-dimensional radial fashion on their own column mass or optical
depth scale this does not matter.  In this sense the Unruh plage model
does recognize that the local temperature gradient around 
\rev{local} $\tau\is1$ 
within magnetic concentrations tends to be less steep than
in the granulation, as in the bottom panels of
Fig.\,\ref{fig:muram-profiles}.  However, the approximation breaks
down for non-vertical fluxtube viewing
%RR radial is confusing for a fluxtube
for which the ``Z\"urich wine-glass'' geometry of
  \citet{1993A&A...268..736B} %T Bunte+Solank+Steiner, wine glass model
with slanted rays passing through the glasses was a much more
realistic \rev{description}, and it fails for limbward faculae because
slanted viewing of \rev{hot} granule innards through empty fluxtubes
(Fig.\,\ref{fig:BPcartoon}) should not be described by the vertical
temperature stratification \revpar within magnetic concentrations.

\rev{Plage and faculae were} much better treated by the older
Solanki-style fluxtube models which diverge with depth instead of with
height between magnetic and non-magnetic (compare the \rev{PLA and
Unruh plage models in Fig.\,\ref{fig:onedimmodels}
and PLA with the MURaM stratification in 
Fig.\,\ref{fig:muram-profiles})}.  Obviously, outward divergence
supplies a zero-order approximation to increasing facular contrast in
limbward viewing, but no more than that and
\rev{inherently wrong}. 

The same criticisms apply to the similar
\rev{photospheric feature modeling through outward-diverging temperature}
stratifications by \eg\
  Fontenla et al.\ (\rev{1993,} 2006),
    \nocite{1993ApJ...406..319F} % FAL including FALC model
    \nocite{2006ApJ...639..441F} %C Fontenla++ refereed by me
%
%RR also FALC, already wrong in upper photosphere
who add more ad-hoc adjustment parameters in the form of a 
\rev{deep-seated} 
chromospheric temperature rise,
\rev{comparable to the ones invoked by
  \citet{2001A&A...369L..13D}}, %C Doyle++ solar MnI variation explained
that was rightfully removed by
  \citet{1999A&A...345..635U} %C Unruh++ irradiance modeling
in \revpar their plage model --
%% to avoid unrealistic self-reversed line cores.
%RR but that is because of the unrealistic assumption of LTE
\rev{the chromosphere has nothing to do with network and 
plage visibility in photospheric diagnostics.}

Nevertheless, the success of the SATIRE modeling by
  \citet{2007msfa.conf..189D} %C Danilovic++ magnetic source MnI
implies that, given any trick to make a single magnetic concentration
brighter in \Mnline\ than in \Feline\ by \revpar the amount given by
SATIRE's two-model divergence, that trick will reproduce Livingston's
data similarly.
\rev{Thus, although we have not performed any full-disk 
modeling, our trick is likely to reproduce these data too.}  
\revpar Our trick entails
better understanding of why \MnI\ lines activity-brighten more than
other lines: \revpar the latter brighten more in normal granulation.

%% Livingston SiI line??   @ Nikola 
%RR either explain or don't mention at all, also in intro.
%RR Neckel profile doesn't look boxy

%%%%%%%%%%%%%%%%%%%%%%%%%%%%%%%%%%%%%%%%%%%%%%%%%%%%%%%%%%%%%%%%%%%%%%%%%%%%
\section{Conclusion}  %RR thought up in April biking Apache Point
%%%%%%%%%%%%%%%%%%%%%%%%%%%%%%%%%%%%%%%%%%%%%%%%%%%%%%%%%%%%%%%%%%%%%%%%%%%%
\revpar
The explanation of the activity sensitivity of \Mnline\ concerns
deep-photosphere line formation only.  Intergranular magnetic
concentrations brighten with respect to field-free intergranular lanes
in any \rev{photospheric} diagnostic through deep radiation escape \revpar
sampling relatively high and flat-gradient temperatures
(Figs.~\ref{fig:muram-slices-A} and \ref{fig:muram-slices-B}).
\rev{For normal, narrow photospheric lines this brightening has less
effect in full-disk averaging through their} loss of line depth in 
\revpar normal granulation (Fig.\,\ref{fig:muram-profiles}) that was
traditionally mimicked by applying micro- and macroturbulent smearing
(Fig.\,\ref{fig:onedimdemo}). The flat-bottomed profile which \Mnline\
possesses thanks to its hyperfine structure
(Fig.\,\ref{fig:onedimdemo}) makes this line much less susceptible to
granular Doppler smearing and thermal broadening \rev{so that it}
weakens less \revpar in \rev{normal} granulation
(Fig.\,\ref{fig:muram-profiles}) and so displays larger \rev{mean}
brightness contrast between quiet and magnetic areas
(Fig.\,\ref{fig:muram-images}).
\revpar
%%In retrospect, the initial paper by
%%%
%%  \citet{1978SoPh...59..275E} %C Elste+Teske: sensitivity network
%%%
%%which triggered the \MnI--versus--activity saga already contained this
%%conclusion by pointing out than \MnI\ hyperfine broadening overrides
%%microturbulence. \revpar 
%% In our opinion this remains the best paper in this saga.

\revpar
The \Mnline\ line is therefore an unsigned proxy magnetometer
sensitive \revpar to the magnetic concentrations that constitute
on-disk \rev{network and plage} and near-limb faculae, through
Wilson-depression viewing of subsurface bright-wall heat near disk
center and through slanted facular viewing into hot granules near the
limb (Fig.\,\ref{fig:BPcartoon}).  In solar irradiance monitoring
\Mnline\ tracks well with the \CaIIHK\ and \MgIIhk\ core brightnesses
because these also respond to magnetic concentrations, although
through unidentified magnetic chromosphere heating that does not
affect the \MnI\ lines, neither directly nor through interlocking to
\MgII\ \hk.
%RR straws of course

\revpar
As a proxy magnetometer \Mnline\ is similar to the G band in which the
contrast enhancement arises from the general addition of CH line
opacity and local reduction of that through dissociation in magnetic
concentrations, and to the extended blue wings of strong lines in
which the contrast enhancement arises from the general addition of
line opacity with reduction of that through lesser damping in magnetic
\rev{concentrations} plus Doppler flattening of the granular contrast
  (\cite{2006A&A...449.1209L}). %C Leenaarts++ Ha BPs
These are all sufficiently wide in wavelength to not suffer from the
granular Doppler smearing that spoils the contrast for the centers of
narrow lines.  The G band is the most useful of these proxies by being
a wide-band spectral feature in the blue.  Qua contrast, the blue wing
of \Halpha\ is probably the best of all
  (\cite{2006A&A...452L..15L}), %C Leenaarts++ Ha best BP 
but not for full-disk irradiance monitoring since its photospheric
magnetic-concentration brightening is sometimes obscured by overlying dark
blue-shifted and/or heat-widened chromospheric fibril absorption.  

We conclude that the principal usefulness of photospheric \MnI\ lines
lies not in their unusual activity sensitivity but in their
hyperfine-structured richness as weak-field diagnostic in full-Stokes
polarimetry with high angular resolution and sensitivity
 (\eg\ \cite{2008ApJ...675..906S}). %C Sanchez++: QS B measurement from HFS
%

%%%%%%%%%%%%%%%%%%%%%%%%%%%%%%%%%%%%%%%%%%%%%%%%%%%%%%%%%% ACKNOWLEDGEMENTS
\begin{acknowledgements}
  We thank J.~S\'anchez Almeida for bringing us together and
  H.~Uitenbroek for valuable interpretation \rev{and many text improvements}.
  N.~Vitas is indebted to I.~Vince for suggesting this research topic.
  His research is supported by a Marie Curie Early Stage Research
  Training Fellowship of the EC's Sixth Framework Programme under
  contract number MEST-CT-2005-020395.  B.~Viticchi\`e thanks V.~Penza
  for invaluable help and J.~S\'anchez Almeida for introducing him to
  \MnI\ lines.  His research is supported by a Regione Lazio CVS
  (Centro per lo studio della variabilit\`a del Sole) PhD grant.
  R.J.~Rutten thanks the National Solar Observatory/Sacramento Peak
  for much hospitality and the Leids Kerkhoven-Bosscha Fonds for
  travel support.
\end{acknowledgements}

%RR R.J.~Rutten's research is not supported by any of his three affiliations.

%%%%%%%%%%%%%%%%%%%%%%%%%%%%%%%%%%%%%%%%%%%%%%%%%%%%%%%%%%%%%%%% REFERENCES
%%\bibliographystyle{aa}
%%\bibliography{aajour,/tmp/rjrfiles.bib,/tmp/adsfiles.bib}

\begin{thebibliography}{72}
\expandafter\ifx\csname natexlab\endcsname\relax\def\natexlab#1{#1}\fi

\bibitem[{{Abt}(1952)}]{1952ApJ...115..199A}
{Abt}, A. 1952, \apj, 115, 199

\bibitem[{{Asensio Ramos} {et~al.}(2007){Asensio Ramos}, {Mart{\'{\i}}nez
  Gonz{\'a}lez}, {L{\'o}pez Ariste}, {Trujillo Bueno}, \&
  {Collados}}]{2007ApJ...659..829A} {Asensio Ramos}, A.,
  {Mart{\'{\i}}nez Gonz{\'a}lez}, M.~J., {L{\'o}pez Ariste}, A.,
  {Trujillo Bueno}, J., \& {Collados}, M. 2007, \apj, 659, 829

\bibitem[{{Auer} {et~al.}(1977){Auer}, {House}, \&
  {Heasley}}]{1977SoPh...55...47A} {Auer}, L.~H., {House}, L.~L., \&
  {Heasley}, J.~N. 1977, \solphys, 55, 47

\bibitem[{{Bergemann} \& {Gehren}(2007)}]{2007A&A...473..291B}
{Bergemann}, M. \& {Gehren}, T. 2007, \aap, 473, 291

\bibitem[{{Blackwell-Whitehead} {et~al.}(2005){Blackwell-Whitehead},
  {Pickering}, {Pearse}, \& {Nave}}]{2005ApJS..157..402B}
  {Blackwell-Whitehead}, R.~J., {Pickering}, J.~C., {Pearse}, O., \&
  {Nave}, G.  2005, \apjs, 157, 402

\bibitem[{{Bruls} \& {Solanki}(1993)}]{1993A&A...273..293B}
{Bruls}, J.~H.~M.~J. \& {Solanki}, S.~K. 1993, \aap, 273, 293

\bibitem[{{B{\"{u}}nte} {et~al.}(1993){B{\"{u}}nte}, {Solanki}, \&
  {Steiner}}]{1993A&A...268..736B} {B{\"{u}}nte}, M., {Solanki},
  S.~K., \& {Steiner}, O. 1993, \aap, 268, 736

\bibitem[{{Canfield}(1971)}]{1971A&A....10...64C}
{Canfield}, R.~C. 1971, \aap, 10, 64

\bibitem[{{Carlsson} {et~al.}(2004){Carlsson}, {Stein}, {Nordlund}, \&
  {Scharmer}}]{2004ApJ...610L.137C} {Carlsson}, M., {Stein}, R.~F.,
  {Nordlund}, {\AA}., \& {Scharmer}, G.~B. 2004, \apjl, 610, L137

\bibitem[{{Cram} {et~al.}(1980){Cram}, {Rutten}, \&
  {Lites}}]{1980ApJ...241..374C} {Cram}, L.~E., {Rutten}, R.~J. \&
  {Lites}, B.~W. 1980, \apj, 241, 374
%RR wrong order in ADS

\bibitem[{{Danilovi{\'c}} {et~al.}(2007){Danilovi{\'c}}, {Solanki},
  {Livingston}, {Krivova}, \& {Vince}}]{2007msfa.conf..189D}
  {Danilovi{\'c}}, S., {Solanki}, S.~K., {Livingston}, W., {Krivova},
  N., \& {Vince}, I. 2007, in Modern solar facilities,
  ed.  F.~{Kneer}, K.~G. {Puschmann}, \& A.~D. {Wittmann},
  189

\bibitem[{{Danilovic} \& {Vince}(2004)}]{2004SerAJ.169...47D}
{Danilovic}, S. \& {Vince}, I. 2004, Serbian Astron.\ Journal, 169, 47
%##

\bibitem[{{Danilovic} \& {Vince}(2005)}]{2005MmSAI..76..949D}
{Danilovic}, S. \& {Vince}, I. 2005, Memorie della Soc.\ Astron.\
  Italiana, 76, 949
%##

\bibitem[{{Danilovic} {et~al.}(2005){Danilovic}, {Vince}, {Vitas}, \&
  {Jovanovic}}]{2005SerAJ.170...79D}
{Danilovic}, S., {Vince}, I., {Vitas}, N., \& {Jovanovic}, P. 2005, Serbian
  Astron.\ Journal, 170, 79
%## 

\bibitem[{{De Wijn} {et~al.}(2005){De Wijn}, {Rutten}, {Haverkamp}, \&
  {S{\"u}tterlin}}]{2005A&A...441.1183D} {De Wijn}, A.~G., {Rutten},
  R.~J., {Haverkamp}, E.~M.~W.~P., \& {S{\"u}tterlin}, P. 2005, \aap,
  441, 1183

\bibitem[{{Doyle} {et~al.}(2001){Doyle}, {Jevremovi{\'c}}, {Short},
  {Hauschildt}, {Livingston}, \& {Vince}}]{2001A&A...369L..13D}
  {Doyle}, J.~G., {Jevremovi{\'c}}, D., {Short}, C.~I., {et~al.} 2001,
  \aap, 369, L13

\bibitem[{{Elste}(1987)}]{1987SoPh..107...47E}
{Elste}, G. 1987, \solphys, 107, 47

\bibitem[{{Elste} \& {Teske}(1978)}]{1978SoPh...59..275E}
{Elste}, G. \& {Teske}, R.~G. 1978, \solphys, 59, 275

\bibitem[{{Fligge} {et~al.}(2000){Fligge}, {Solanki}, \&
  {Unruh}}]{2000A&A...353..380F} {Fligge}, M., {Solanki}, S.~K., \&
  {Unruh}, Y.~C. 2000, \aap, 353, 380

\bibitem[{{Fontenla} {et~al.}(2006){Fontenla}, {Avrett}, {Thuillier}, \&
  {Harder}}]{2006ApJ...639..441F} {Fontenla}, J.~M., {Avrett}, E.,
  {Thuillier}, G., \& {Harder}, J. 2006, \apj, 639, 441

\bibitem[{{Fontenla} {et~al.}(1993){Fontenla}, {Avrett}, \&
  {Loeser}}]{1993ApJ...406..319F} {Fontenla}, J.~M., {Avrett}, E.~H.,
  \& {Loeser}, R. 1993, \apj, 406, 319

\bibitem[{{Gurtovenko} \& {Kostyk}(1989)}]{1989KiIND.........G}
{Gurtovenko}, E.~A. \& {Kostyk}, R.~I. 1989, Naukova Dumka, Kiev
%##

\bibitem[{{Keller} {et~al.}(2004){Keller}, {Sch{\"u}ssler}, {V{\"o}gler}, \&
  {Zakharov}}]{2004ApJ...607L..59K} {Keller}, C.~U., {Sch{\"u}ssler},
  M., {V{\"o}gler}, A., \& {Zakharov}, V. 2004, \apjl, 607, L59

\bibitem[{{Krivova} {et~al.}(2003){Krivova}, {Solanki}, {Fligge}, \&
  {Unruh}}]{2003A&A...399L...1K} {Krivova}, N.~A., {Solanki}, S.~K.,
  {Fligge}, M., \& {Unruh}, Y.~C. 2003, \aap, 399, L1

\bibitem[{{Kurucz}(1979)}]{1979ApJS...40....1K}
{Kurucz}, R.~L. 1979, \apjs, 40, 1

\bibitem[{{Kurucz}(1992{\natexlab{a}})}]{1992RMxAA..23..181K}
{Kurucz}, R.~L. 1992{\natexlab{a}}, Revista Mexicana Astron.\
  Astrofis., 23, 181
%##

\bibitem[{{Kurucz}(1992{\natexlab{b}})}]{1992RMxAA..23..187K}
{Kurucz}, R.~L. 1992{\natexlab{b}}, Revista Mexicana Astron.\ 
  Astrofis., 23, 187
%##

\bibitem[{{Landi degl'Innocenti}(1978)}]{1978A&AS...33..157L}
{Landi degl'Innocenti}, E. 1978, \aaps, 33, 157

\bibitem[{{Leenaarts} {et~al.}(2006{\natexlab{a}}){Leenaarts}, {Rutten},
  {Carlsson}, \& {Uitenbroek}}]{2006A&A...452L..15L} {Leenaarts}, J.,
  {Rutten}, R.~J., {Carlsson}, M., \& {Uitenbroek}, H.
  2006{\natexlab{a}}, \aap, 452, L15

\bibitem[{{Leenaarts} {et~al.}(2006{\natexlab{b}}){Leenaarts}, {Rutten},
  {S{\"u}tterlin}, {Carlsson}, \& {Uitenbroek}}]{2006A&A...449.1209L}
  {Leenaarts}, J., {Rutten}, R.~J., {S{\"u}tterlin}, P., {Carlsson},
  M., \& {Uitenbroek}, H. 2006{\natexlab{b}}, \aap, 449, 1209

\bibitem[{{Leenaarts} {et~al.}(2005){Leenaarts}, {S{\"u}tterlin}, {Rutten},
  {Carlsson}, \& {Uitenbroek}}]{2005ESASP.596E..15L} {Leenaarts}, J.,
  {S{\"u}tterlin}, P., {Rutten}, R.~J., {Carlsson}, M., \&
  {Uitenbroek}, H. 2005, in Chromospheric and Coronal Magnetic Fields,
  ed. D.~E. {Innes}, A.~{Lagg}, \& S.~A. {Solanki}, ESA Special Pub.\
  596, 15
%##

\bibitem[{{Lemaire} \& {Skumanich}(1973)}]{1973A&A....22...61L}
{Lemaire}, P. \& {Skumanich}, A. 1973, \aap, 22, 61

\bibitem[{{Livingston} \& {Wallace}(1987)}]{1987ApJ...314..808L}
{Livingston}, W. \& {Wallace}, L. 1987, \apj, 314, 808

\bibitem[{{Livingston} {et~al.}(2007){Livingston}, {Wallace}, {White}, \&
  {Giampapa}}]{2007ApJ...657.1137L} {Livingston}, W., {Wallace}, L.,
  {White}, O.~R., \& {Giampapa}, M.~S. 2007, \apj, 657, 1137

\bibitem[{{L{\'o}pez Ariste} {et~al.}(2006{\natexlab{a}}){L{\'o}pez Ariste},
  {Ram{\'{\i}}rez V{\'e}lez}, {Tomczyk}, {Casini}, \&
  {Semel}}]{2006ASPC..358...54L} {L{\'o}pez Ariste}, A.,
  {Ram{\'{\i}}rez V{\'e}lez}, J.~C., {Tomczyk}, S., {Casini}, R., \&
  {Semel}, M. 2006{\natexlab{a}}, in Astron.\ Soc.\ Pacific Conf.\
  Series, 358,  %%## ed. R.~{Casini} \& B.~W. {Lites}, 
  54
%## no booktitle

\bibitem[{{L{\'o}pez Ariste} {et~al.}(2002){L{\'o}pez Ariste}, {Tomczyk}, \&
  {Casini}}]{2002ApJ...580..519L} {L{\'o}pez Ariste}, A., {Tomczyk},
  S., \& {Casini}, R. 2002, \apj, 580, 519

\bibitem[{{L{\'o}pez Ariste} {et~al.}(2006{\natexlab{b}}){L{\'o}pez Ariste},
  {Tomczyk}, \& {Casini}}]{2006A&A...454..663L} {L{\'o}pez Ariste},
  A., {Tomczyk}, S., \& {Casini}, R. 2006{\natexlab{b}}, \aap, 454,
  663

\bibitem[{{Malanushenko} {et~al.}(2004){Malanushenko}, {Jones}, \&
  {Livingston}}]{2004IAUS..223..645M} {Malanushenko}, O., {Jones},
  H.~P., \& {Livingston}, W. 2004, in Multi-Wavelength Investigations
  of Solar Activity, ed. A.~V.  {Stepanov}, E.~E. {Benevolenskaya}, \&
  A.~G. {Kosovichev}, IAU Symp.\ 223, 645
%##

\bibitem[{{Maltby} {et~al.}(1986){Maltby}, {Avrett}, {Carlsson},
  {Kjeldseth-Moe}, {Kurucz}, \& {Loeser}}]{1986ApJ...306..284M}
  {Maltby}, P., {Avrett}, E.~H., {Carlsson}, M., {et~al.} 1986, \apj,
  306, 284

\bibitem[{{Milkey} \& {Mihalas}(1974)}]{1974ApJ...192..769M}
{Milkey}, R.~W. \& {Mihalas}, D. 1974, \apj, 192, 769

\bibitem[{Neckel(1999)}]{Neckel1999}
Neckel, H. 1999, Sol.\ Phys., 184, 421

\bibitem[{{Orozco Su{\'a}rez} {et~al.}(2007){Orozco Su{\'a}rez}, {Bellot
  Rubio}, {del Toro Iniesta}, {Tsuneta}, {Lites}, {Ichimoto},
  {Katsukawa}, {Nagata}, {Shimizu}, {Shine}, {Suematsu}, {Tarbell}, \&
  {Title}}]{2007ApJ...670L..61O} {Orozco Su{\'a}rez}, D., {Bellot
  Rubio}, L.~R., {del Toro Iniesta}, J.~C., {et~al.} 2007, \apjl, 670,
  L61

\bibitem[{Rutten(1988)}]{Rutten1988b}
Rutten, R.~J. 1988, in Physics of Formation of FeII Lines Outside LTE, ed.
  R.~Viotti, A.~Vittone, \& M.~Friedjung, IAU Coll.\ 94,
  185
%##

\bibitem[{{Rutten}(1999)}]{1999ASPC..184..181R}
{Rutten}, R.~J. 1999, in Magnetic Fields and Oscillations,
ed. B.~{Schmieder}, A.~{Hofmann}, \& J.~{Staude}, Astron.\
Soc. Pacific Conf.\ Series, 184, 181
%##

\bibitem[{Rutten(2003)}]{Rutten2003:RTSA}
Rutten, R.~J. 2003, Radiative Transfer in Stellar Atmospheres,
  Lecture Notes Utrecht University,
  \url{http://www.astro.uu.nl/~rutten} 
%##

\bibitem[{{Rutten} \& {Kostik}(1982)}]{1982A&A...115..104R}
{Rutten}, R.~J. \& {Kostik}, R.~I. 1982, \aap, 115, 104

\bibitem[{{Rutten} \& {Stencel}(1980)}]{1980A&AS...39..415R}
{Rutten}, R.~J. \& {Stencel}, R.~E. 1980, \aaps, 39, 415

\bibitem[{{Rutten} \& {van der Zalm}(1984)}]{1984A&AS...55..143R}
{Rutten}, R.~J. \& {van der Zalm}, E.~B.~J. 1984, \aaps, 55, 143

\bibitem[{{S{\'a}nchez Almeida} {et~al.}(2008){S{\'a}nchez Almeida},
  {Viticchi{\'e}}, {Landi Degl'Innocenti}, \&
  {Berrilli}}]{2008ApJ...675..906S} {S{\'a}nchez Almeida}, J.,
  {Viticchi{\'e}}, B., {Landi Degl'Innocenti}, E., \& {Berrilli},
  F. 2008, \apj, 675, 906

\bibitem[{{Shelyag} {et~al.}(2004){Shelyag}, {Sch{\"u}ssler}, {Solanki},
  {Berdyugina}, \& {V{\"o}gler}}]{2004A&A...427..335S} {Shelyag}, S.,
  {Sch{\"u}ssler}, M., {Solanki}, S.~K., {Berdyugina}, S.~V., \&
  {V{\"o}gler}, A. 2004, \aap, 427, 335

\bibitem[{{Sheminova} {et~al.}(2005){Sheminova}, {Rutten}, \& {Rouppe van der
  Voort}}]{2005A&A...437.1069S} {Sheminova}, V.~A., {Rutten}, R.~J.,
  \& {Rouppe van der Voort}, L.~H.~M. 2005, \aap, 437, 1069

\bibitem[{{Skumanich} \& {Lites}(1987)}]{1987ApJ...322..473S}
{Skumanich}, A. \& {Lites}, B.~W. 1987, \apj, 322, 473

\bibitem[{{Solanki}(1986)}]{1986A&A...168..311S}
{Solanki}, S.~K. 1986, \aap, 168, 311

\bibitem[{{Solanki} \& {Brigljevic}(1992)}]{1992A&A...262L..29S}
{Solanki}, S.~K. \& {Brigljevic}, V. 1992, \aap, 262, L29

\bibitem[{{Solanki} \& {Steenbock}(1988)}]{1988A&A...189..243S}
{Solanki}, S.~K. \& {Steenbock}, W. 1988, \aap, 189, 243

\bibitem[{{Spruit}(1976)}]{1976SoPh...50..269S}
{Spruit}, H.~C. 1976, \solphys, 50, 269

\bibitem[{{Staath} \& {Lemaire}(1995)}]{1995A&A...295..517S}
{Staath}, E. \& {Lemaire}, P. 1995, \aap, 295, 517

\bibitem[{{Thackeray}(1937)}]{1937ApJ....86..499T}
{Thackeray}, A.~D. 1937, \apj, 86, 499

\bibitem[{{Unruh} {et~al.}(1999){Unruh}, {Solanki}, \&
  {Fligge}}]{1999A&A...345..635U} {Unruh}, Y.~C., {Solanki}, S.~K., \&
  {Fligge}, M. 1999, \aap, 345, 635

\bibitem[{{Vince} \& {Erkapic}(1998)}]{1998IAUS..185..459V}
{Vince}, I. \& {Erkapic}, S. 1998, in New
Eyes to See Inside the Sun and Stars, ed. F.-L. {Deubner},
J.~{Christensen-Dalsgaard}, \& D.~{Kurtz}, IAU Symp.\ 185, 459
%##

\bibitem[{{Vince} {et~al.}(2005{\natexlab{a}}){Vince}, {Gopasyuk}, {Gopasyuk},
  \& {Vince}}]{2005SerAJ.170..115V} {Vince}, I., {Gopasyuk}, O.,
  {Gopasyuk}, S., \& {Vince}, O. 2005{\natexlab{a}}, Serbian
  Astron.\ Journal, 170, 115
%##

\bibitem[{{Vince} {et~al.}(2005{\natexlab{b}}){Vince}, {Vince}, {Ludm{\'a}ny},
  \& {Andriyenko}}]{2005SoPh..229..273V} {Vince}, I., {Vince}, O.,
  {Ludm{\'a}ny}, A., \& {Andriyenko}, O.  2005{\natexlab{b}},
  \solphys, 229, 273

\bibitem[{{Vitas}(2005)}]{2005MSAIS...7..164V}
{Vitas}, N. 2005, Memorie Soc.\ Astron.\ Italiana Suppl. 7, 164
%##

\bibitem[{{Vitas} \& {Vince}(2007)}]{2007ASPC..368..543V}
{Vitas}, N. \& {Vince}, I. 2007, in The Physics of Chromospheric
Plasmas, ed.  P.~{Heinzel}, I.~{Dorotovi{\v c}}, \& R.~J. {Rutten},
Astron.\ Soc.\ Pacific Conf.\ Series, 368, 543
%##

\bibitem[{{V{\"o}gler}(2004)}]{2004A&A...421..755V}
{V{\"o}gler}, A. 2004, \aap, 421, 755

\bibitem[{{V{\"o}gler} \& {Sch{\"u}ssler}(2003)}]{2003AN....324..399V}
{V{\"o}gler}, A. \& {Sch{\"u}ssler}, M. 2003, Astron.\
Nachrichten, 324, 399
%##

\bibitem[{{V{\"o}gler} {et~al.}(2005){V{\"o}gler}, {Shelyag}, 
  {Sch{\"u}ssler}, {Cattaneo}, {Emonet}, \&
  {Linde}}]{2005A&A...429..335V} {V{\"o}gler}, A., {Shelyag}, S.,
  {Sch{\"u}ssler}, M., {et~al.} 2005, \aap, 429, 335

\bibitem[{Wallace {et~al.}(1998)Wallace, Hinkle, \&
  Livingston}]{Wallace+Hinkle+Livingston1998} Wallace, L., Hinkle, K.,
  \& Livingston, W. 1998, An Atlas of the Spectrum of the Solar
  Photosphere from 13,500 to 28,000~cm$^{-1}$ (3570 to 7405~\AA),
  Technical Report 98-001, National Solar Observatory, Tucson
%##

\bibitem[{{Wenzler} {et~al.}(2005){Wenzler}, {Solanki}, \&
  {Krivova}}]{2005A&A...432.1057W} {Wenzler}, T., {Solanki}, S.~K., \&
  {Krivova}, N.~A. 2005, \aap, 432, 1057

\bibitem[{{Wenzler} {et~al.}(2006){Wenzler}, {Solanki}, {Krivova}, \&
  {Fr{\"o}hlich}}]{2006A&A...460..583W} {Wenzler}, T., {Solanki},
  S.~K., {Krivova}, N.~A., \& {Fr{\"o}hlich}, C. 2006, \aap, 460, 583

\bibitem[{{Westendorp Plaza} {et~al.}(1998){Westendorp Plaza}, {del Toro
  Iniesta}, {Ruiz Cobo}, {Mart{\'{\i}}nez Pillet}, {Lites}, \&
  {Skumanich}}]{1998ApJ...494..453W} {Westendorp Plaza}, C., {del Toro
  Iniesta}, J.~C., {Ruiz Cobo}, B., {et~al.}  1998, \apj, 494, 453

\bibitem[{{Zwaan}(1967)}]{1967SoPh....1..478Z}
{Zwaan}, C. 1967, \solphys, 1, 478

\end{thebibliography}
%% file: mni-clean.bbl = cleaned up for final insertion
%% init: Jul 13 2008 
%% last: Jul 13 2008 
%% note: %## = edited the item above

%RR cleaned up bbl - now permanent!

%%%%%%%%%%%%%%%%%%%%%%%%%%%%%%%%%%%%%%%%%%%%%%%%%%%%%%%%%%%%%%%%%%%%%%% END
\end{document}